%% file: main.tex
\documentclass[a4paper,12pt,twoside]{article}

\usepackage[T1]{fontenc}
\usepackage{lmodern}
\usepackage[english]{babel}
\usepackage{csquotes}

\usepackage{amsmath,amsthm,amssymb}
\usepackage{tensor}
\usepackage{tikz-cd}

\usepackage[compat=1.1.0]{tikz-feynman}
\usepackage[margin=2.54cm]{geometry} 
\usepackage{setspace}
\usepackage{titling}
\pretitle{\begin{center} \LARGE \bfseries }
\usepackage{caption}
\usepackage{subcaption}
\usepackage[affil-it]{authblk}

\usepackage{multirow}
\usepackage{xcolor}
\usepackage[normalem]{ulem}

\usepackage[backend=biber,style=phys,biblabel=brackets,sorting=none,eprint=true]{biblatex}
\usepackage{hyperref}

\DeclareMathOperator*{\Res}{Res}
\newcommand{\minus}{\scalebox{0.5}[.8]{$ - $}}

\title{On-shell recursion and soft theorems for worldline quantum field theory}
\author{Nathan Castet}
\author{Vincent F. He}
\affil{Institut des Hautes Études Scientifiques, 91440 Bures-sur-Yvette, France}
\date{}

\begin{document}

% --- Front page ---
\maketitle
\thispagestyle{empty}
\begin{abstract}
    We derive a set of on-shell recursion relations for amplitudes in worldline quantum field theory (WQFT), an effective theory of gravity coupled to a massive worldline which is relevant for computing  classical gravitational observables.
    We overcome the poor large-$z$ falloff of amplitudes under an all-line complex momentum deformation by deriving and exploiting a new soft-graviton theorem in the presence of a worldline.
    This theorem generalizes the leading (Weinberg) and subleading soft theorems with additional contributions due to soft emission from the worldline.
    Our main result is a $d$-dimensional recursion formula, expressing any $n$-graviton rational amplitude in terms of lower-point amplitudes from WQFT and pure gravity.
    Using the soft-worldline theorem, we can obtain amplitudes with any number of external worldline fluctuations from the corresponding all-graviton amplitudes, making the recursion sufficient to construct any rational amplitude of the theory.
    We verify our results by explicitly computing two-graviton and three-graviton amplitudes.
\end{abstract}

\newpage

% --- Table of content ---
\tableofcontents
\pagenumbering{arabic}

\newpage

% --- Body ---
\input{sections/01-introduction}
\input{sections/03-worldline-qft}
\input{sections/04-soft-theorems}
\input{sections/05-recursion-algorithm}
\input{sections/06-results}
\input{sections/07-conclusion}

%\newpage

\section*{Acknowledgments}

    We are deeply grateful to Julio Parra-Martinez for his guidance, constant support, and insightful discussions throughout the course of this work. 
    We also thank him for providing invaluable advice in the preparation of this manuscript.
    V.F.H. acknowledges support from the Simons Foundation.
    The work of N.C. was supported by the European Research Council (ERC) under the European Union's Horizon 2020 research and innovation programme, grant agreement ERC-AdG-885414 Ampl2Einstein, PI David A. Kosower, and under the European Union's Horizon Europe research and innovation programme, grant agreement 101221094 (ERC Starting Grant GravitaS), PI Julio Parra-Martinez.
    Views and opinions expressed are those of the authors only and do not necessarily reflect those of the European Union or the European Research Council. Neither the European Union nor the granting authority can be held responsible for them.

%\newpage

% --- Appendix ---
\appendix
\input{appendices/a-soft-z-theorems}
\input{appendices/b-local-gauge-invariant-tensor-decomposition}
\input{appendices/c-spinor-helicity}

\printbibliography

\end{document}

%% file: sections/01-introduction.tex
\section{Introduction}

    Over the last decade, experimental success in gravitational-waves detection, led by the LIGO-Virgo-KAGRA collaboration has laid the foundation for a new era of precision gravitational astrophysics \cite{Abbott2016,Abbott2017,Abbott2023}.
    As tremendous momentum is building behind third-generation detectors, expected substantial improvements in sensitivity fuel the need for high-precision theoretical modelling of physical observables \cite{Purrer2020}.

    Thanks to remarkable efforts, a variety of theoretical approaches have been developed  over the recent years~\cite{Pretorius2005,Lehner2014,Mino1997,Barack2018,Blanchet2024,Buonanno1999,Damour2009}.
    Applicable to both scattering black holes and to the initial inspiral phase of binary systems, the post-Minkowskian (PM) framework notably leverages quantum field theory (QFT) techniques to obtain classical observables as an expansion in the gravitational constant $G$, valid at all orders in velocity~\cite{Kosower2018,Bern2019,Kalin2020,Bern2022,Driesse2025,Dlapa2021,Buonanno2022,Sagex13,Sagex14}.
    In that spirit, worldline quantum field theory (WQFT), by describing compact objects as point-particles on a worldline~
    \cite{Goldberger2006,Plefka2021}, has proven to be efficient for computing classical observables at high PM order~\cite{Driesse2024,Jakobsen2023,Haddad2025}.
    In particular, the subtle limiting procedure of previous amplitude-based methods for extracting classical observables from quantum-gravity amplitudes is made particularly simple in WQFT. 

    A strategy going back to Berends and Giele~\cite{Berends1987}, constructs scattering amplitudes recursively from off-shell currents and has recently been imported into the worldline setting~\cite{Jakobsen2023,Hoogeveen2025}.
    While powerful, the method remains intrinsically off-shell: the recursion is seeded by an off-shell expression rather than by on-shell lower-point amplitudes.
    The gauge ambiguities inherent to off-shell objects still require careful handling.
    On the other hand, on-shell methods proved to be an extremely valuable tool in computing gravitational amplitudes, bypassing the use of cumbersome Feynman rules for a spin-2 field.
    It is therefore desirable to have a genuinely on-shell recursion for worldline amplitudes, assembled directly from on-shell lower-point data.

    Such fully on-shell recursions are famously illustrated by the work of Britto, Cachazo, Feng and Witten (BCFW)~\cite{Britto2005a,Britto2005b}.
    Given a scattering amplitude, one defines a deformed version of the external kinematics, parameterized by a complex variable $z$.
    The shifted amplitude $A(z)$ is a rational function, with poles that can be traced back to deformed internal propagators going on-shell.
    Using an appropriate contour integral, it is then possible to relate the unshifted amplitude to the residues of this new function.
    As a general consequence of unitarity, these residues factorize into lower-point amplitudes.
    When the contour integral at infinity vanishes, the relation takes a particularly clean form: the original amplitude is completely determined by simpler amplitudes.
    This is guaranteed to happen if $A(z)$ decays fast enough at large $z$.
    For a wide range of physically interesting theories however, poor large-$z$ behavior makes finding such a deformation difficult.
    These include EFTs with higher-dimensional operators and in particular WQFT --- the main focus of this article.

    %Achieving this for worldline amplitudes requires taming their large-$z$ behavior.
    Indeed, as we will later explain, an amplitude with $m$ external worldline fluctuations scales as $z^m$.
    This leads to an undetermined boundary contribution, which obstructs a naive application of basic recursion methods.
    The present work builds on existing on-shell methods for WQFT~\cite{He2025}, to derive a new set of recursion formulas, valid in arbitrary spacetime dimension.
    We overcome the poor large-$z$ behavior by drawing on the strategy described in~\cite{Cheung2016,Luo2016}; 
    we introduce soft poles through modified contour integrals, whose residues are fixed by a new soft graviton theorem in the presence of a worldline, which we derive in the paper.
    This systematically improves the large-$z$ falloff and eliminates these boundary terms.
    
    The paper is organized as follows.
    In section \ref{sec:worldline-qft} we provide a brief exposition of WQFT.
    Section \ref{sec:soft-theorems} is dedicated to the infrared structure of the theory. 
    In particular we give a new theorem for soft gravitons in the presence of a worldline, which we derived from Feynman rules and gauge invariance.
    Section \ref{sec:recursion} contains the main result of the paper: we display an all-line soft recursion for $n \geq 3$ graviton amplitudes, which in WQFT are related to the classical part of generalized Compton amplitudes.
    We then treat the case $n = 2$ separately.
    We checked our explicit results for the two- and three-graviton amplitudes against independent computations.
    We present the four-dimensional results in section \ref{sec:results} using spinor-helicity variables for compactness. 
    The full $d$-dimensional results are included in an auxiliary Mathematica file.

    \paragraph{Conventions:}
    We use the mostly-minus metric $\eta_{\mu \nu} = \operatorname{diag}(+,-,-,\dots)$ in $d$-dimension and we take the gravitational coupling $\kappa^2 = 32 \pi G$.
    External momenta of amplitudes are taken to be outgoing.
    Generic WQFT amplitudes with $n$ external graviton and $m$ worldline fluctuation are denoted $A_{g^n w^m}$.
    When the amplitude has external gravitons only ($m=0$), we often use the lighter notation $A_n$.
    For pure gravity amplitudes (i.e. derived from the Einstein-Hilbert part of the action only) we write $A^{\text{EH}}_n$.

%% file: sections/03-worldline-qft.tex
\section{WQFT : A review}\label{sec:worldline-qft}

    In this section, we present a short introduction to aspects of WQFT that are relevant for the purposes of this work.
    A more in-depth exposition and a comparison with the traditional scalar-graviton QFT is given in \cite{Plefka2021}.

    A massive, non-spinning compact object is described at long distances by an EFT of a spinless point-particle with worldline action
    \begin{equation}\label{eq:wl-action}
        S_\text{wl}[g,x] = - \frac{m}{2} \int d \tau g_{\mu \nu} \Dot{x}^\mu \Dot{x}^\nu + \dots,
    \end{equation}
    where the dots denote higher-dimensional operators on the worldline, encoding the internal structure of the compact body.
    In this work we restrict to the minimal action, setting the Wilson coefficients of these operators to zero. 
    Their contributions are suppressed by powers of $R\,\omega$, with $R$ the size of the body and $\omega$ the frequency of the gravitons, and for compact objects with $R \sim Gm$ they only appear at higher order in $G$.
    %These do not contribute up to the perturbative order that we consider in this paper and we can safely ignore them \cite{Plefka2021}.
    
    %At higher order some contact terms may not be fixed purely by gauge invariance and correspond to genuinely new operators to be fixed.

    Hence the full action of the theory with a single massive particle has the form 
    \begin{equation}\label{eq:action}
        S[g,x] = S_\text{EH}[g] + S_\text{gf}[g] + S_\text{wl}[g,x],
    \end{equation}
    where 
    \begin{equation}
        S_\text{EH}[g] = -\frac{2}{\kappa} \int d^d x \sqrt{-g} R
    \end{equation}
    denotes the Einstein-Hilbert action and $S_\text{gf}$  the gauge fixing term, which we can leave unspecified since our on-shell recursion will be gauge invariant.
    Of course this can be easily generalized to multiple worldlines by replacing $S_\text{wl}[g,x]$ with a sum of terms $\sum_i S_\text{wl}[g,x_i]$.
    The usual way to proceed is to expand the fields $g_{\mu \nu}$ and $x^\mu$ around background values corresponding to a flat metric and a straight trajectory
    \begin{equation}
        g_{\mu \nu} = \eta_{\mu \nu} + \kappa h_{\mu \nu}, \qquad x^\mu(\tau) =   b^\mu + \tau u^\mu + \delta x^\mu(\tau),
    \end{equation}
    where the four-velocity satisfies $u^2=1$, $b$ is the impact parameter and we refer to $\delta x$ as the worldline fluctuation.
    Note that the background value of $x^\mu$ breaks the full translation symmetry of Minkowski spacetime to translation invariance in the $u^\mu$ direction.
    As a result, energy-momentum conservation is relaxed to energy conservation in the $u^\mu$ frame (i.e. energy in the frame corresponding to the background trajectory of the worldline).
    Hence instead of having the usual momentum preserving delta functions $\delta^{(d)}(\sum_i k_i)$ in front of our amplitudes we will find $\delta(\sum_i u\cdot k_i)$, where $k_i$ are the momenta carried away from the worldline by gravitons.

    Let us now briefly review the Feynman rules in WQFT. (A detailed derivation and discussion can be found in \cite{Plefka2021}.) Expanding $S_{\text{wl}}$ around the background values of the fields, we get the free field equation of motion $\delta \Ddot{x}^\mu = 0$, which implies the on-shell condition $\omega = 0$ for the energy carried by a worldline fluctuation.
    The propagators of the theory are
    \begin{equation}\label{eq:propagators}
        \vcenter{\hbox{\input{diagrams/z-propagator}}} = \frac{-i \eta_{\rho \sigma}}{m \omega^2},
        \qquad
        \vcenter{\hbox{\input{diagrams/g-propagator}}} = \frac{i P_{\mu \nu ; \alpha \beta}}{k^2}.
    \end{equation}
    On the left we have the worldline fluctuation and we recognize the usual graviton propagator on the right. In de Donder gauge the projection tensor is $P_{\mu \nu ; \alpha \beta} = \frac{1}{2}(\eta_{\mu \alpha} \eta_{\nu \beta} + \eta_{\mu \beta} \eta_{\nu \alpha} - \frac{2}{d-2}\eta_{\mu \nu} \eta_{\alpha \beta})$.
    These propagators must be accompanied by an appropriate choice of $i\epsilon$ prescription. 
    Since the rational amplitudes considered in this work (see Sec.~\ref{sec:AGO}) involve no integration, all prescriptions, whether time-symmetric or retarded, yield the same rational function away from its poles, and the recursion derived below holds for any of them.
    The prescription only becomes relevant for amplitudes that involve integration.

    The $n$-graviton vertices arise from the Einstein-Hilbert part of the action and are the same as in pure gravity, an expression for the three-graviton vertex can be found in \cite{Donoghue1994}.
    In particular, these vertices obey the full energy-momentum conservation.

    The worldline part of the action gives an infinite tower of vertices with one external graviton and any number of external fluctuations.
    The one and two-point vertices are
    \begin{equation}\label{eq:vert0}
        \vcenter{\hbox{\input{diagrams/g-vertex}}} = -\frac{i m \kappa}{2} e^{i k \cdot b}\Bar{\delta} (u \cdot k) u^\mu u^\nu,
    \end{equation}
    \begin{equation}\label{eq:vert1}
        \vcenter{\hbox{\input{diagrams/gz-vertex}}} = \frac{m \kappa}{2} e^{i k \cdot b}\Bar{\delta} (u \cdot k + \omega) \big(2 \omega u^{(\mu} \delta^{\nu)}_\rho + u^\mu u^\nu k_\rho \big),
    \end{equation}
    where we used the modified delta distribution $\Bar{\delta} = 2 \pi \delta$.
    The dotted lines are drawn as a visual guide.
    They do not correspond to any dynamical field or particle in the theory but rather represent the background worldline as a source.

    \subsection{Amplitudes and gravitational observables}\label{sec:AGO}

        A variety of classical observables can be expressed in this framework.
        For example in the presence of two worldlines, one can describe the radiated gravitational field from the scattering of two black holes as a specific amplitude
        \begin{equation}
            \lim_{k^2 \to 0} k^2 \langle h_{\mu \nu}(k) \rangle = \; \input{diagrams/waveform},
        \end{equation}
        where the angle brackets represent correlation functions.
        For the same scattering event, the impulse on either black hole is written as
        \begin{equation}
            \Delta p_i^\mu = - m_i \lim_{\omega^2 \to 0} \omega^2 \langle \delta x_i^\mu(\omega) \rangle = \; \input{diagrams/impulse}.
        \end{equation}
        From a QFT perspective both of these amplitudes are tadpoles; the dotted lines represent the background trajectories of the two worldlines, which behave as classical sources for the $h_{\mu \nu}$ and $\delta x^\mu$ fields.
        The prefactor $k^2$ (or $\omega^2$) implements the usual LSZ amputation of the external propagator, while the limit places the corresponding field on shell.
        In this sense, these classical observables inherit the same on-shell, amputated structure as a standard scattering amplitude.

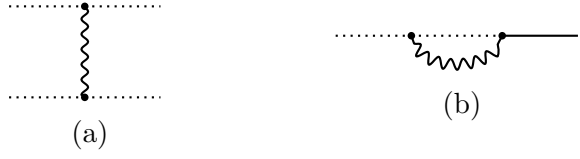
\begin{figure}[t]
            \centering
            \begin{subfigure}[]{0.3\textwidth}
                \centering
                \input{diagrams/integral-diagram-a}
                \caption{}
                \label{fig:non-rational-trees-a}
            \end{subfigure}
            \begin{subfigure}[]{0.3\textwidth}
                \centering
                \input{diagrams/integral-diagram-b}
                \caption{}
                \label{fig:non-rational-trees-b}
            \end{subfigure}
                \caption{Examples of tree diagrams that involve an integral and do not give a rational amplitude in the worldline theory. The worldline vertices in Eqs. \eqref{eq:vert0} and \eqref{eq:vert1} only constrain the energy component of the graviton; some of the components remain unconstrained and must be integrated over. The integral in (b) is scaleless so it vanishes in dimreg but may give non-zero a contribution with another regulation scheme.}
            \label{fig:non-rational-trees}
        \end{figure}

        Interestingly, the classical quantities are obtained by only keeping the tree level diagrams.
        This is to be contrasted with other PM techniques based on interaction with massive scalar fields, where the classical-quantum separation is more subtle.
        Nevertheless, this does not exempt us from having to evaluate Feynman integrals. Amplitudes that do not require integration are rational functions of the external kinematics.
        In a typical QFT, these exactly coincide with tree-level amplitudes. 
        However in WQFT, rational amplitudes only constitute a subset of tree-level amplitudes.
        Indeed, because worldline vertices do not impose the full momentum conservation, the presence of multiple sources can leave some integral over the transferred momenta between them.
        Figure~\ref{fig:non-rational-trees} depicts basic examples of such tree-level diagrams involving an integral.
        A useful characterization is that a diagram is rational if and only if shrinking all worldlines to a single point yields a tree diagram.
        Note that in these amplitudes, the only dependence on the impact parameter $b$ is through a global phase factor $e^{i b \cdot k_{\text{tot}}}$, where $k_{\text{tot}} = \sum_i k_i$ is the sum of all the external momenta. 
        We therefore call the amplitudes rational, disregarding this exponential factor.

        In the presence of a single worldline, the two-graviton amplitude is closely related to the gravitational Compton amplitude of a massive scalar, written in terms of $\bar p = (p+p')/2 = m u$, where $p$ ($p'$) is the incoming (outgoing) momentum of the scalar.
        At leading order in $G$, the WQFT amplitude below corresponds to the leading term in the large-mass expansion at fixed graviton momenta~\cite{Plefka2021,Brandhuber2021}.
        Physically, it describes the scattering of a gravitational wave off a black hole.
        % In the presence of a single worldline, amplitudes with two external gravitons are related to the usual gravitational Compton process.
        % Classically this corresponds to gravitational wave scattering on a black hole.
        %The Feynman rules straightforwardly give
        At leading order, Eq.~\eqref{eq:2gAmp} receives two diagrammatic contributions, and Eq.~\eqref{eq:2gwAmp} receives three.
        For an explicit computation see~\cite{BjerrumBohr2025}.
        We find
        \begin{equation}\label{eq:2gAmp}
            A_{g^2}(k_1,k_2) \quad = \quad \vcenter{\hbox{\input{diagrams/compton}}} \quad = \quad - \frac{ \mathcal{K}_{12}^2}{4 (k_1 \! \cdot \! k_2)(u \! \cdot \! k_1)^2},
        \end{equation}
        \begin{multline}\label{eq:2gwAmp}
            A_{g^2 w}(k_1,k_2,\omega) \quad = \quad \input{diagrams/g2w} \\
            = -i \zeta_1 \! \cdot (k_1 + k_2) \frac{ \mathcal{K}_{12}^2}{4 (k_1 \! \cdot \! k_2)(u \! \cdot \! k_1)^2}
            - i \omega\frac{\mathcal{K}_{12}}{2 (k_1 \! \cdot \! k_2)(u \! \cdot \! k_1)^3} \big[(k_1 \! \cdot \! k_2)(u \! \cdot \! e_1)(u \! \cdot \! e_2)(k_1 \! \cdot \! \zeta_1)  \\[.5ex]
            + (u \! \cdot \! k_1)^2 ((e_1 \! \cdot \! k_2)(e_2 \! \cdot \! \zeta_1) -(e_2 \! \cdot \! k_1)(e_1 \! \cdot \! \zeta_1) + (e_1 \! \cdot \! e_2)(k_1 \! \cdot \! \zeta_1)) \\
            - (k_1 \! \cdot \! k_2)(u \! \cdot \! k_1) ( (e_1 \! \cdot \! \zeta_1)(u \! \cdot \! e_2) + (e_2 \! \cdot \! \zeta_1)(u \! \cdot \! e_1)) \big], \raisetag{5em}
        \end{multline}
        with the kinematic factor\footnote{Introducing the linearized field strengths $F_i^{\mu\nu} = k_i^\mu e_i^\nu - e_i^\mu k_i^\nu$, energy conservation $u\cdot k_2 = -u\cdot k_1$ gives $\mathcal K_{12} = u\cdot F_1\cdot F_2\cdot u$, making gauge invariance manifest and Eq.~\eqref{eq:2gAmp} recognizable as the double copy of the heavy-mass Compton amplitude~\cite{Brandhuber2021}. In Eq.~\eqref{eq:2gwAmp}, this identity does not hold away from $\omega = 0$, so the form is kept as in~\eqref{eq:k_factor}.}
        \begin{equation}\label{eq:k_factor}
            \mathcal{K}_{12}
            = u \! \cdot \! k_1\left[(e_1 \! \cdot \! k_2)(u \! \cdot \! e_2) + (e_1 \! \cdot \! e_2)(u \! \cdot \! k_1)\right] - u \! \cdot \! e_1\left[(k_1 \! \cdot \! k_2)(u \! \cdot \! e_2) + (e_2 \! \cdot \! k_1)(u \! \cdot \! k_1)\right].
        \end{equation}
        Here $e_i$ and $\zeta_i$ denote the corresponding graviton and worldline fluctuation polarizations, respectively.
        In Eq.~\eqref{eq:2gwAmp} we have kept the $\mathcal{O}(\omega)$ part that one would find when computing the amplitude from the Feynman rules, before setting $\omega = 0$. 
        Also, in both cases we omit a global factor of $m \kappa^2$.

        Keeping track of this part which is linear in $\omega$ turns out to be important for using on-shell methods like factorization.
        Indeed, an amplitude is singular and factorizes when an internal propagator is nearly on-shell
        \begin{equation}
            A(\omega_{\text{int}}) = \frac{A_{\mathrm{L}}(\omega_{\text{int}}) A_{\mathrm{R}}(-\omega_{\text{int}})}{\omega_{\text{int}}^2} + \mathcal{O}(\omega_{\text{int}}^0).
        \end{equation}
        Take $A_{\mathrm{L,R}}(\omega_{\text{int}})$ to be defined through the Feynman rules, where every external leg has been put on-shell except for the worldline fluctuation with energy $\omega_{\text{int}}$.
        As this formula suggests, we expect our amplitude to have both a double and a simple pole.
        This can be seen in Eq. \eqref{eq:2gAmp}, with $\omega_{\text{int}} = u \! \cdot \! k_1$.
        Now clearly, if we wanted to determine the singular structure of our amplitude with this method, we would need to know the $\mathcal{O}(\omega_{\text{int}})$ parts of $A_{\mathrm{L}}$ and $A_{\mathrm{R}}$.
        A systematic on-shell approach to constructing rational amplitudes and integrands --- including the linear-in-$\omega$ parts --- was recently provided by \cite{He2025}.
        The method relies on a complexification procedure $\omega^2 \to \omega \Bar{\omega}$. 
        In this same article the authors show how the $\mathcal{O}(\omega)$ part of an amplitude is in fact a gauge invariant quantity that can be obtained from the corresponding amplitude without the external worldline fluctuation.
        We will review some of these results in Sec.~\ref{sec:soft-theorems}.

    \subsection{Large-\texorpdfstring{$z$}{z} behavior of WQFT amplitudes}
    
        The recursion relations we will derive below will rely on an all-line complex deformation such as $k_i \to k_i(1-z/z_i)$, and require control over the large-$z$ behaviour of WQFT amplitudes  $A_{g^n w^m}(z)$. 
        Because we consider an all-line shift, the large-$z$ scaling power is simply given by dimensional analysis (this is, accounting for the mass dimension of the momenta only, ignoring coupling constants such as $\kappa$ and masses of the point-particles).
        For $n=1$ and $m=0,1$ the mass dimension can be read directly from the worldline vertices~\eqref{eq:vert0} and~\eqref{eq:vert1}.
        More generally, the $m$-th order worldline vertex as provided in~\cite{Plefka2021} has a mass dimension of $m$.
        We now generalize this result to amplitudes with an arbitrary number of external gravitons and worldline fluctuations.

        Starting with the base case: an amplitude with external gravitons only ($m=0$) has mass dimension zero.
        This can be seen by induction; any diagram contributing to the $n$-graviton amplitude can be obtained by appending a graviton line to an ($n{-}1$)-graviton diagram.
        If the line connects to a graviton vertex, the dimension of the diagram is unchanged as every pure gravity vertex has dimension 2.
        If this new leg connects to a graviton propagator, this procedure introduces a three-graviton vertex (dimension 2) and an extra graviton propagator (dimension $-2$), again leaving the global dimension unaltered. 
        The same conclusion holds if the extra graviton is attached to a worldline fluctuation line.
        Since the one-graviton vertex has dimension zero, the same is true for any amplitude with $m=0$.

        Extending this logic, a diagram contributing to $A_{g^n w^m}$ can be obtained by attaching an external worldline fluctuation to an existing worldline vertex, increasing the total dimension by one.
        By induction, this confirms that the mass dimension --- and thus the large-$z$ scaling power --- is~$m$.
        As a check, one can easily see that the amplitudes above have the expected large-$z$ behavior.

        This power counting relies on using the minimal action~\eqref{eq:action}. 
        Higher-dimensional worldline operators come with Wilson coefficients of negative mass dimension, and their contributions grow faster at large $z$. For instance, the tidal operator $c_E\int d\tau\,E_{\mu\nu}E^{\mu\nu}$ contributes to the two-graviton amplitude a local term of the schematic form
        \begin{equation}
            A^{\rm tidal}_{g^2}(k_1,k_2) \sim c_E\,\kappa^2 K_{12}^2 ,
        \end{equation}
        which has no poles and scales as $z^4$, in contrast with Eq.~\eqref{eq:2gAmp}.
        Similar remarks apply to higher-curvature corrections to the bulk action, such as $R^3$ terms or, for $d>4$, the Gauss--Bonnet term, which would modify the pure-gravity amplitudes $A^{\rm EH}$ entering the recursion and worsen the large-$z$ behaviour. 
        We do not consider them here.

%% file: diagrams/z-propagator.tex
\begin{tikzpicture}[baseline={(current bounding box.center)}]
\begin{feynman}

    % Graviton legs
    \vertex (u1) at (-.8,0);
    \vertex (v1) at (0,0);
    \node [circle, fill=black, inner sep=1pt] at (v1) {};
    \vertex (v2) at (1,0);
    \node [circle, fill=black, inner sep=1pt] at (v2) {};
    \vertex (u2) at (1.8,0);
    
    \draw[thick, dotted] (u1) -- (v1);
    \draw[thick] (v1) -- (v2);
    \draw[thick, dotted] (v2) -- (u2);
    
    % Labels
    \node at (-.1,0.3) {\small $\rho$};
    \node at (1.1,0.3) {\small $\sigma$};
    \node at (.5,-.3) {$\omega$};

\end{feynman}
\end{tikzpicture}

%% file: diagrams/g-propagator.tex
\begin{tikzpicture}[baseline={(current bounding box.center)}]
\begin{feynman}

    % Graviton legs
    \vertex (v1) at (0,0);
    \node [circle, fill=black, inner sep=1pt] at (v1) {};
    \vertex (v2) at (1,0);
    \node [circle, fill=black, inner sep=1pt] at (v2) {};
    
    \draw[thick,photon] (v1) -- (v2);
    
    % Labels
    \node at (-.1,0.3) {\small $\mu \nu$};
    \node at (1.1,0.3) {\small $\alpha \beta$};
    \node at (.5,-.3) {$k$};

\end{feynman}
\end{tikzpicture}

%% file: diagrams/g-vertex.tex
\begin{tikzpicture}[baseline={(current bounding box.center)}]
\begin{feynman}

    % Graviton legs
    \vertex (u1) at (-.8,0);
    \vertex (v1) at (0,0);
    \node [circle, fill=black, inner sep=1pt] at (v1) {};
    \vertex (u2) at (.8,0);
    \vertex (v2) at (0,-.8);
    
    \draw[thick, dotted] (u1) -- (v1);
    \draw[thick, dotted] (v1) -- (u2);
    \draw[thick, photon] (v1) -- (v2);
    
    % Labels
    \node at (.35,-.8) {\small $\mu \nu$};

\end{feynman}
\end{tikzpicture}

%% file: diagrams/gz-vertex.tex
\begin{tikzpicture}[baseline={(current bounding box.center)}]
\begin{feynman}

    % Graviton legs
    \vertex (u1) at (-.8,0);
    \vertex (v1) at (0,0);
    \node [circle, fill=black, inner sep=1pt] at (v1) {};
    \vertex (u2) at (.8,0);
    \vertex (v2) at (0,-.8);
    
    \draw[thick, dotted] (u1) -- (v1);
    \draw[thick] (v1) -- (u2);
    \draw[thick, photon] (v1) -- (v2);
    
    % Labels
    \node at (.8,.25) {\small $\rho$};
    \node at (.35,-.8) {\small $\mu \nu$};

\end{feynman}
\end{tikzpicture}

%% file: diagrams/waveform.tex
\begin{tikzpicture}[baseline={([yshift=1.5ex]current bounding box.center)}]
    \begin{feynman}

        % Worldline through blob
        \vertex (wtl) at (-1, .35);
        \vertex (wtr) at (1, .35);
        \draw[thick, dotted] (wtl) -- (wtr);
        
        \vertex (wbl) at (-1, -.35);
        \vertex (wbr) at (1, -.35);
        \draw[thick, dotted] (wbl) -- (wbr);

        % Blob
        \node[draw, circle, fill=gray!30, minimum size=1cm] (blob) {};

        % Graviton leg (bottom)
        \vertex (k) at (.8, -.8);
        \draw[thick, photon] (blob.280) -- (k);

        % Labels
        \node at (1, -1) {$k$};

    \end{feynman}
\end{tikzpicture}

%% file: diagrams/impulse.tex
\begin{tikzpicture}[baseline={([yshift=-.5ex]current bounding box.center)}]
    \begin{feynman}
    
        % Worldline through blob
        \vertex (wtl) at (-1, .35);
        \vertex (wtr) at (1, .35);
        \vertex (v) at (0, .35);
        \draw[thick, dotted] (wtl) -- (v);
        \draw[thick] (v) -- (wtr);
        
        \vertex (wbl) at (-1, -.35);
        \vertex (wbr) at (1, -.35);
        \draw[thick, dotted] (wbl) -- (wbr);
        
        % Blob
        \node[draw, circle, fill=gray!30, minimum size=1cm] (blob) {};
        
        % Labels
        \node at (1.3, .35) {$\omega$};
        
    \end{feynman}
\end{tikzpicture}

%% file: diagrams/integral-diagram-a.tex
\begin{tikzpicture}[baseline={(current bounding box.center)}]
\begin{feynman}

    % Graviton legs
    \vertex (u1) at (-1,0);
    \vertex (v1) at (0,0);
    \node [circle, fill=black, inner sep=1pt] at (v1) {};
    \vertex (u2) at (1,0);
    \vertex (u3) at (-1,-1.2);
    \vertex (v2) at (0,-1.2);
    \node [circle, fill=black, inner sep=1pt] at (v2) {};
    \vertex (u4) at (1,-1.2);

    \draw[thick, dotted] (u1) -- (v1);
    \draw[thick, dotted] (v1) -- (u2);
    \draw[thick, photon] (v1) -- (v2);
    \draw[thick, dotted] (u3) -- (v2);
    \draw[thick, dotted] (v2) -- (u4);

\end{feynman}
\end{tikzpicture}

%% file: diagrams/integral-diagram-b.tex
\begin{tikzpicture}[baseline={(current bounding box.center)}]
\begin{feynman}

    % Graviton legs
    \vertex (u1) at (-1,0);
    \vertex (v1) at (0,0);
    \node [circle,fill=black,inner sep=1pt] at (v1) {};
    \vertex (v2) at (1.2,0);
    \node [circle,fill=black,inner sep=1pt] at (v2) {};
    \vertex (u2) at (2.2,0);
    
    \draw[thick, dotted] (u1) -- (v1);
    \draw[thick, dotted] (v1) -- (v2);
    \draw[decorate, decoration={
        snake, 
        amplitude=2pt, 
        segment length=5pt, 
        pre length=3pt, 
        post length=3pt},
        thick, 
        in=-110, out=-80] (v1) to (v2);
    \draw[thick] (v2) -- (u2);

\end{feynman}
\end{tikzpicture}

%% file: diagrams/compton.tex
\begin{tikzpicture}[baseline={([yshift=6ex]current bounding box.center)}]
    \begin{feynman}
    
        % Worldline through blob
        \vertex (wtl) at (-1, .15);
        \vertex (wtr) at (1, .15);
        \draw[thick, dotted] (wtl) -- (wtr);

        % Blob
        \node[draw, circle, fill=gray!30, minimum size=.7cm] (blob) {};

        % Graviton leg (bottom)
        \vertex (gl) at (-.5, -.9);
        \vertex (gr) at (.5, -.9);
        \draw[thick, photon] (blob.220) -- (gl);
        \draw[thick, photon] (blob.320) -- (gr);
        
        % Labels
        \node at (-.8, -.9) {$k_1$};
        \node at (1, -.9) {$k_2$};
        
    \end{feynman}
\end{tikzpicture}

%% file: diagrams/g2w.tex
\begin{tikzpicture}[baseline={([yshift=-.5ex]current bounding box.center)}]
    \begin{feynman}
    
        % Worldline through blob
        \vertex (wtl) at (-1, .15);
        \vertex (wtr) at (1, .15);
        \vertex (v) at (0, .15);
        \draw[thick, dotted] (wtl) -- (v);
        \draw[thick] (v) -- (wtr);

        % Blob
        \node[draw, circle, fill=gray!30, minimum size=.7cm] (blob) {};

        % Graviton leg (bottom)
        \vertex (gl) at (-.5, -.9);
        \vertex (gr) at (.5, -.9);
        \draw[thick, photon] (blob.220) -- (gl);
        \draw[thick, photon] (blob.320) -- (gr);

        % Labels
        \node at (1.3, .35) {$\omega$};
        \node at (-.8, -.9) {$k_1$};
        \node at (1, -.9) {$k_2$};
        
    \end{feynman}
\end{tikzpicture}

%% file: sections/04-soft-theorems.tex
\section{Soft theorems}\label{sec:soft-theorems}
    
    In this section we describe soft theorems for amplitudes in WQFT.
    We start by reviewing the leading and subleading theorems for a soft worldline perturbation \cite{He2025}. 
    Then we show that Feynman rules and gauge invariance of gravitons lead to a new soft-graviton theorem.

    \subsection{Soft-worldline theorems}

        Both the leading and subleading soft-worldline theorems follow from spurionic symmetries of the worldline action --- symmetries involving a transformation of the classical background.
        These generate operator equations whose overlaps with on-shell states yield the desired soft behavior.
        Since the derivation relies only on symmetries of the action and the LSZ formula, the two results are non-perturbative.
        As derived, they are statements about in-out amplitudes. 
        The Ward identities follow from a standard path integral, which generates time-ordered correlators.
        For the rational amplitudes considered in this work, the $i \epsilon$ prescription does not matter (see below Eq.~\eqref{eq:propagators}).
        In this section we simply state the theorems,
        see App.~\ref{app:soft_z_derivations} for a derivation.

        As worldline fluctuations are on-shell when $\omega = 0$, it may seem odd to discuss a soft theorem.
        In fact such language follows \cite{He2025}, where the double pole structure of worldline propagators motivated the definition of a complexified worldline energy $\omega^2 \rightarrow \omega \bar{\omega}$.
        The on-shell condition can then be satisfied by setting $\bar{\omega} = 0$ only.
        In that context, an on-shell amplitude with an external worldline fluctuation remains a function of $\omega$, and one can discuss their soft behavior as $\omega \to 0$.
        Although the complexification itself is not necessary for our recursion, we do need the gauge invariant information contained in the $\mathcal{O}(\omega)$ piece of the amplitude to fix the worldline factorizations.
        Thus both results are useful to us, and we retain the name soft theorem.

        \paragraph{Leading soft-worldline theorem.}
        
            The worldline Lagrangian is invariant under the simultaneous shift $b^\mu \to b^\mu + a^\mu$, $\delta x^\mu(\tau) \to \delta x^\mu(\tau) - a^\mu$ for any constant vector $a$. Exploiting this symmetry via an LSZ reduction argument, one finds\footnote{Energy conservation allows one to eliminate one Lorentz invariant product $(u \cdot k_i)$ from the amplitude. In practice, for the soft theorems to hold, the same Lorentz product must be eliminated from both sides for the theorem to apply.}
            \begin{equation}\label{eq:leading_soft_z}
                \lim_{\omega \to 0} A(\omega, \alpha) = i\zeta \! \cdot k_{\text{tot}} \, A(\alpha),
            \end{equation}
            where $k_{\text{tot}} = \sum_i k_i$ is the total external momentum of the hard amplitude $A(\alpha)$ (without the extra worldline fluctuation). 
            This result has a simple interpretation: the amplitude with one additional on-shell worldline is entirely determined by the hard amplitude without it, up to a universal kinematic prefactor.\footnote{Alternatively if the worldline fluctuation is off-shell ($\omega \neq 0$), the leading order in $\omega$ is determined.}
            In particular, it provides a simple algebraic formula for computing an on-shell amplitude with external worldline fluctuations, from the corresponding all-graviton amplitude.
            
            As an illustration we have written the amplitudes in Eqs. \eqref{eq:2gAmp} and \eqref{eq:2gwAmp} in a way that makes this property explicit. Indeed, one might easily check that
            \begin{equation}
                A_{g^2 w}(k_1,k_2,\omega) = i \zeta_1 \! \cdot (k_1 + k_2) A_{g^2}(k_1,k_2) + \mathcal{O}(\omega).
            \end{equation}

        \paragraph{Subleading soft-worldline theorem.}
        
            A second spurionic symmetry, defined by the transformation $u^\mu \to u^\mu + a^\mu$, $\delta x^\mu(\tau) \to \delta x^\mu(\tau) - a^\mu\tau$, yields a constraint on the next order in the soft expansion.
            The derivation, again detailed in App.~\ref{app:soft_z_derivations}, gives
            \begin{equation}
                \lim_{\omega \to 0} \frac{\partial}{\partial \omega} A(\omega, \alpha) 
                = i\zeta^\mu \frac{\partial}{\partial u^\mu} A(\alpha).
            \label{eq:subleading-soft-z}
            \end{equation}
            The right-hand side is evaluated on the amplitude as given by the Feynman rules, and $u^2=1$ can be imposed only after taking the derivative\footnote{The Feynman rules are valid for arbitrary $u$. Setting $u^2=1$ is a gauge choice that sets the affine parameter of the worldline to be the proper time. This prescription is necessary as $\zeta \cdot \partial_u$ depends on how the amplitude is extended away from $u^2=1$. For instance, $A$ and $u^2A$ agree at $u^2=1$, but their derivatives differ by $2(\zeta\cdot u)A$.}.
            Naively the left hand side appears to be off-shell: the partial derivative with respect to $\omega$ is taken while $\omega$ is still off-shell.
            However because the right hand side is on-shell and universal, this is a valid on-shell quantity.
            This theorem plays an essential role in the recursion of Sec.~\ref{sec:recursion}: the double poles arising from worldline-propagators require knowledge of $\partial_\omega A$, which is precisely the information supplied by Eq.~\eqref{eq:subleading-soft-z}.
            The expression can again be checked explicitly when $\alpha$ corresponds to two external gravitons by comparing Eqs.~\eqref{eq:2gAmp} and \eqref{eq:2gwAmp}, although it is not as immediate as the leading result.

            These results are closely related to properties of the worldline vertices discussed in the literature.
            In~\cite{Plefka2021} it was already noted that the vertex with $n+1$ worldline fluctuations satisfies
            \begin{equation}
                V_{\rho_1 \dots \rho_{n+1}} \xrightarrow{\omega_{n+1} \to 0} \frac{\partial}{\partial b^{\rho_{n+1}}} V_{\rho_1 \dots \rho_{n}}.
            \end{equation}
            Since the propagators do not depend on $b$, applying this to every worldline vertex of a rational amplitude reproduces Eq.~\eqref{eq:leading_soft_z} in perturbation theory.
            Some further similarities were described by Eq.~(4.15) of~\cite{Haddad2024} in the form of a Poisson bracket containing also the subleading term, and by Eq.~(3.42) of~\cite{Jakobsen2023PhD} for the vertices of a supersymmetric spinning worldline.

    \subsection{New soft-graviton theorem from gauge invariance}\label{sec:soft-graviton}

        Our recursion will also require an understanding of soft-graviton limits.
        The leading and subleading contributions in the expansion of a $(n{+}1)$-graviton WQFT amplitude in such limit are written
        % \begin{multline}\label{eq:soft-graviton-theorem}
        %     A_{n+1}(q; k_1 \dots k_n) = \frac{\kappa}{2} \bigg\{ \sum_{i=1}^n \frac{e_{\mu \nu}}{k_i q}\bigg(k_i^\mu - \frac{k_i q}{u q}u^\mu \bigg)\bigg(k_i^\nu - \frac{k_i q}{u q}u^\nu \bigg) \\ 
        %     \qquad \qquad \qquad \qquad \qquad \qquad - i \sum_{i=1}^n \frac{e_{\mu \nu}k_i^\mu q_\rho J_i^{\nu \rho}}{k_i q} - i \frac{e_{\mu \nu}u^\mu q_\rho K^{\nu \rho}}{u q} \bigg\} A_{n} (k_1 \dots k_n) \\
        %     + \mathcal{O}(q),
        % \end{multline}
        % \begin{multline}\label{eq:soft-graviton-theorem}
        %     A_{n+1}(q; k_1 \dots k_n) = \frac{\kappa}{2} \bigg\{ \sum_{i=1}^n \frac{1}{k_i q}\left(e k_i - \frac{k_i q}{u q}(u e) \right)^2 \\ 
        %     \qquad \qquad \qquad \qquad \qquad \qquad - i \sum_{i=1}^n \frac{(e k_i) e_{\nu}  q_\rho J_i^{\nu \rho}}{k_i q} - i \frac{(u e) e_{\nu} q_\rho K^{\nu \rho}}{u q} \bigg\} A_{n} (k_1 \dots k_n) \\
        %     + \mathcal{O}(q),
        % \end{multline}
        \begin{multline}\label{eq:soft-graviton-theorem}
            A_{n+1}(q; k_1 \dots k_n) = \frac{\kappa}{2} \bigg\{ \sum_{i=1}^n \frac{1}{k_i \! \cdot \! q}\left(e \! \cdot \! k_i - \frac{k_i \! \cdot \!  q}{u \! \cdot \! q} u \! \cdot \! e \right)^2 \\ 
            \qquad \qquad \qquad \qquad \qquad \qquad - i \sum_{i=1}^n \frac{e \! \cdot \! k_i (e \! \cdot \!  J_i \! \cdot \! q)}{k_i \! \cdot \! q} - i \frac{u \! \cdot \! e (e \! \cdot \! K \! \cdot \! q)}{u \! \cdot \! q} \bigg\} A_{n} (k_1 \dots k_n) \\
            + \mathcal{O}(q),
        \end{multline}
        where $q$ is the soft momentum and $e$ is the polarization of the soft graviton.
        We have defined angular momentum operators $J_i^{\nu \rho} = L_i^{\nu \rho} + \Sigma_i^{\nu \rho}$ for external gravitons with
        \begin{equation}
            L_i^{\nu \rho} = i \bigg( k_i^\nu \frac{\partial}{\partial k_{i \rho}} - k_i^\rho \frac{\partial}{\partial k_{i \nu}}\bigg), \quad \Sigma_i^{\nu \rho} = i \bigg( e_i^\nu \frac{\partial}{\partial e_{i \rho}} - e_i^\rho \frac{\partial}{\partial e_{i \nu}}\bigg),
        \end{equation}
        and for the worldline
        \begin{equation}
            K^{\nu \rho} = i \bigg( u^\nu \frac{\partial}{\partial u_\rho} - u^\rho \frac{\partial}{\partial u_\nu} \bigg).
        \end{equation}
        These differential operators act on the amplitude as a function of unconstrained $k_i$, $e_i$ and $u$. 
        Since $J_i$ and $K$ generate Lorentz rotations of $(k_i, e_i)$ and of $u$ respectively, they are tangent to the on-shell conditions $k_i^2 = e_i\cdot e_i = e_i\cdot k_i = 0$ and to $u^2=1$ (e.g.\ $K u^2 = 0$, in contrast with $\zeta\cdot\partial_u$ in Eq.~\eqref{eq:subleading-soft-z}).
        Hence the result does not depend on how the amplitude is extended off these constraints and needs no prescription.
    
        Equation~\eqref{eq:soft-graviton-theorem} is new, and generalizes the leading and subleading Weinberg soft theorems from pure gravity~\cite{Weinberg1965,Cachazo2014,Bern2014} by incorporating worldline-specific contributions through the velocity $u^\mu$ and the worldline angular momentum operator $K^{\nu \rho}$.
        %The first term of each line on the right hand side of \eqref{eq:soft-graviton-theorem} exactly reproduces the leading and subleading Weinberg soft factors that are familiar from pure gravity \cite{Weinberg1965,Cachazo2014,Bern2014}, while the remaining terms are worldline-specific contributions.
        Our formula holds for rational amplitudes and there could be some loop corrections at higher orders.
        We now derive this result.

        \paragraph{Derivation.} 
        It was shown in \cite{Bern2014} that using gauge invariance is a very efficient way to obtain subleading behaviors of tree amplitudes in gravitational theories.
        We follow a similar procedure to obtain our soft theorem for an amplitude with multiple external gravitons in WQFT.
        Looking at the Feynman diagrams, it is clear that diagrams where the soft graviton connects to an internal line will be regular in the soft limit.
        Therefore only the contributions where the soft graviton connects to an external line will have a pole as $q \to 0$.

        \begin{figure}[t]
            \centering
            \begin{subfigure}[]{0.4\textwidth}
                \centering
                \input{diagrams/np1-graviton-singular-a}
                \caption{}
                \label{fig:softGravitonExtLeg}
            \end{subfigure}
            \begin{subfigure}[]{0.4\textwidth}
                \centering
                \input{diagrams/np1-graviton-singular-b}
                \caption{}
                \label{fig:softGravitonGZ}
            \end{subfigure}
            \caption{Singular contributions to the $(n{+}1)$-graviton amplitude in the soft limit}
        \end{figure}
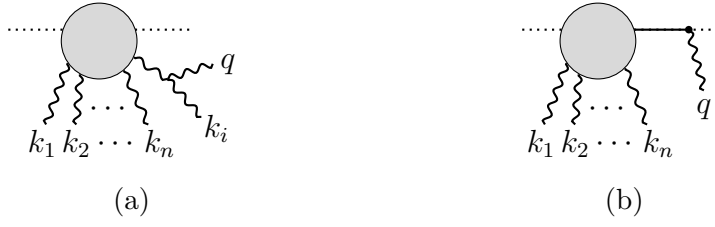
        
        With this in mind, we can organize the terms of the $(n{+}1)$-graviton amplitude in the following way
        \begin{multline}\label{eq:Agnp1} 
            A_{n+1}^{\mu \nu ; \mu_1 \nu_1 \dots \mu_n \nu_n}(q; k_1 \dots k_n) 
            = \frac{\kappa}{2}  \sum_{i=1}^n \frac{1}{k_i \! \cdot \! q}\left[k_i^\mu \eta^{\mu_i \alpha} - i q_\rho (\Sigma_i^{\mu \rho})^{\mu_i \alpha} \right] \left[k_i^\nu \eta^{\nu_i \beta} - i q_\rho (\Sigma_i^{\nu \rho})^{\nu_i \beta} \right] \\
            \times {A_{n}^{\mu_1 \nu_1 \dots}}_{\alpha \beta}{}^{\dots \mu_n \nu_n}(k_1 \dots k_i+q \dots k_n) \\
            - i \frac{\kappa}{2} \frac{ u^\mu u^\nu q_\rho-2(u \! \cdot \! q)u\rule{0pt}{1.8ex}^{(\mu} \delta^{\nu)}_\rho}{(u \! \cdot \! q)^2} A_{g^n w}^{\rho ; \mu_1 \nu_1 \dots \mu_n \nu_n}(u \! \cdot \! q; k_1 \dots k_n) \\
            + N^{\mu \nu ; \mu_1 \nu_1 \dots \mu_n \nu_n}(q; k_1 \dots k_n),
        \end{multline}
        with $(\Sigma_i^{\mu \rho})^{\mu_i \alpha} = i(\eta^{\mu \mu_i}\eta^{\alpha \rho} - \eta^{\mu \alpha}\eta^{\mu_i \rho})$.
        The first term corresponds to diagrams where the soft graviton connects to an external graviton, as depicted in Fig.~\ref{fig:softGravitonExtLeg}.
        The second term corresponds to diagrams represented in Fig.~\ref{fig:softGravitonGZ}, where the argument $u\! \cdot \! q$ of $A_{g^n w}$ is the energy of the worldline, while the last term is symmetric in $(\mu \leftrightarrow \nu)$ and groups all the remaining non-singular contributions.
        The simplified three-graviton vertex used in the first line is obtained by (i) keeping only terms that do not vanish when the amplitude is contracted with the polarizations, and (ii) absorbing all terms that have a factor of $k_i \! \cdot \! q$, cancelling the singularity inside of $N^{\mu \nu ; \mu_1 \nu_1 \dots \mu_n \nu_n}$.
        Gauge invariance requires that the contraction of this amplitude with the soft momentum $q_\mu$ vanishes
        \begin{multline}\label{eq:ward-identity-constraint}
            q_\mu A_{n+1}^{\mu \nu ; \mu_1 \nu_1 \dots \mu_n \nu_n}(q; k_1 \dots k_n) \\
            = \frac{\kappa}{2}  \sum_{i=1}^n \left[k_i^\nu \eta^{\nu_i \beta} - i q_\rho (\Sigma_i^{\nu \rho})^{\nu_i \beta} \right] {A_{n}^{\mu_1 \nu_1 \dots \mu_i}}_{\beta}{}^{\dots \mu_n \nu_n}(k_1 \dots k_i+q \dots k_n) \\
            +i  \frac{\kappa}{2} \delta^\nu_\rho A_{g^n w}^{\rho ; \mu_1 \nu_1 \dots \mu_n \nu_n}(u\! \cdot \!q; k_1 \dots k_n) + q_\mu N^{\mu \nu ; \mu_1 \nu_1 \dots \mu_n \nu_n}(q; k_1 \dots k_n) \\
            = 0.
        \end{multline}
        At order 0 in $q$, this constraint becomes 
        \begin{equation}\label{eq:leadingSoftZ}
            A_{g^n w}^{\rho ; \mu_1 \nu_1 \dots \mu_n \nu_n}(0;k_1,\dots,k_n) = i \sum_{i=1}^n (k_i^\rho) A_{n}^{\mu_1 \nu_1 \dots \mu_n \nu_n}(k_1,\dots,k_n).
        \end{equation}
        Contracting this expression with external polarizations we recognize none other than our leading soft-worldline theorem.
        This is a non-trivial consistency check: the leading soft-worldline theorem, originally derived from a spurionic symmetry of the action, is recovered here as a direct consequence of graviton gauge invariance.
        
        Expanding \eqref{eq:ward-identity-constraint} at linear order in $q$, we find the constraint
        \begin{multline}
            \frac{\kappa}{2} \sum_{i=1}^n \left[ k_i^\nu \eta^{\nu_i \beta} q_\rho \frac{\partial}{\partial k_{i\rho}} - i q_\rho (\Sigma_i^{\nu \rho})^{\nu_{i} \beta} \right]{A_{n}^{\mu_1 \nu_1 \dots \mu_i}}_{ \beta}{}^{\dots \mu_n \nu_n}(k_1 \dots k_n) \\
            + i \frac{\kappa}{2} \delta_\rho^\nu (u \! \cdot \! q) \frac{\partial}{\partial \omega} A_{g^n w}^{\rho ; \mu_1 \nu_1 \dots \mu_n \nu_n}(0;k_1,\dots,k_n) \\
            + q_\mu N^{\mu \nu; \mu_1 \nu_1 \dots \mu_n \nu_n}(0; k_1 \dots k_n) = 0.
        \end{multline}
        So keeping the symmetric part and up to a gauge invariant term $E$ we have 
        \begin{multline}\label{eq:LON}
            N^{\mu \nu ; \mu_1 \nu_1 \dots \mu_n \nu_n}(0; k_1 \dots k_n) = -\frac{1}{2} \sum_{i=1}^n\left( k_i^\mu \frac{\partial}{\partial k_{i \nu}} + k_i^\nu \frac{\partial}{\partial k_{i \mu}} \right) A_{n}^{\mu_1 \nu_1\dots \mu_n \nu_n}(k_1 \dots k_n)  \\
            + u^{(\mu}\delta^{\nu)}_\rho \frac{\partial}{\partial \omega} A_{g^n w}^{\rho ; \mu_1 \nu_1 \dots \mu_n \nu_n}(0; k_1 \dots k_n) + E^{\mu \nu ; \mu_1 \nu_1 \dots \mu_n \nu_n}(k_1 \dots k_n).
        \end{multline}
        The most general rational $E^{\mu \nu}$ satisfying gauge invariance $q_\mu E^{\mu \nu} = q_\nu E^{\mu \nu} = 0$, while also being local in $q$ (i.e. non-singular as $q \rightarrow 0$) can be written as a sum of terms, each of the form
        \begin{equation}
            [(B_1 \! \cdot \! q) B_2^\mu - (B_2 \! \cdot \! q) B_1^\mu][(B_3 \! \cdot \! q) B_4^\nu - (B_4 \! \cdot \! q) B_3^\nu],
        \end{equation}
        where $B_{1,2,3,4}$ are rational, Lorentz vector expressions of the kinematic variables, that are local in $q$ (we give a proof of that claim in Appendix~\ref{app:local-gauge-invariant}).
        As any such term vanishes when $q \rightarrow 0$, there cannot be any other contribution to $N^{\mu \nu ; \mu_1 \nu_1 \dots \mu_n \nu_n}(0; k_1 \dots k_n)$.
        Therefore~\eqref{eq:LON} is already the correct expression, as there is no gauge invariant term that would satisfy all hypotheses while being sufficiently suppressed in $q$.

        Organizing~\eqref{eq:Agnp1} as a $q$-expansion and applying the soft-worldline theorems, we obtain our soft-graviton theorem; contracting with external polarizations then gives the final form given in~\eqref{eq:soft-graviton-theorem}.

        As an immediate check let us specialize to the case $n=1$.
        Using the one-point amplitude
        \begin{equation}\label{eq:g1-wqft}
            A_1(k_1) = -\frac{1}{2}(u \! \cdot \! e_1)^2,
        \end{equation}
        the soft-graviton theorem becomes
        \begin{multline}
            A_2(q; k_1) = \frac{1}{2 k_1 \! \cdot \! q}\left(e \! \cdot \! k_1 - \frac{k_1 \! \cdot \! q}{u \! \cdot \! q}u \! \cdot \! e \right)^2 \left(-\frac{1}{2} (u \! \cdot \! e_1)^2 \right) \\
            + \frac{(u \! \cdot \! e_1)(e \! \cdot \! k_1)((e_1 \! \cdot \! q)(u \! \cdot \! e) - (e_1 \! \cdot \! e)(u \! \cdot \! q))}{2 k_1 \! \cdot \! q} + \frac{(u \! \cdot \! e_1)(u \! \cdot \! e)((e_1 \! \cdot \! e)(u \! \cdot \! q) - (e_1 \! \cdot \! q)(u \! \cdot \! e))}{2 u \! \cdot \! q} \\
            + \mathcal{O}(q).
        \end{multline}
        On the other hand rearranging Eq.~\eqref{eq:2gAmp} as a soft expansion in $k_2$, and using energy conservation $u \! \cdot \! k_1 \to -u \! \cdot \! k_2$ to make the $k_2$ dependence apparent gives
        \begin{multline}\label{eq:A2_soft_expansion}
            A_2(k_1, k_2) = -\frac{1}{4} \frac{(u \! \cdot \! e_1)^2}{k_1 \! \cdot \! k_2}\left(e_2 \! \cdot \! k_1 - \frac{k_1 \! \cdot \! k_2}{u \! \cdot \! k_2}u \! \cdot \! e_2 \right)^2 \\
            + \frac{(u \! \cdot \! e_1)(e_2 \! \cdot \! k_1)((e_1 \! \cdot \! k_2)(u \! \cdot \! e_2) - (e_1 \! \cdot \! e_2)(u \! \cdot \! k_2))}{2 k_1 \! \cdot \! k_2} + \frac{(u \! \cdot \! e_1)(u \! \cdot \! e_2)((e_1 \! \cdot \! e_2)(u \! \cdot \! k_2) - (e_1 \! \cdot \! k_2)(u \! \cdot \! e_2))}{2 u \! \cdot \! k_2} \\
            - \frac{((e_1 \! \cdot \! e_2)(u \! \cdot \! k_2) - (e_1 \! \cdot \! k_2)(u \! \cdot \! e_2))^2}{4 k_1 \! \cdot \! k_2}.
        \end{multline}
        The two expressions clearly agree as $q=k_2$ and $e_2 = e$.

        % Compare this expression to our soft theorem~\eqref{eq:soft-graviton-theorem} and with the one-graviton amplitude
        % \begin{equation}
        %     A_1(k_1) = -\frac{1}{2}(u e_1)^2,
        % \end{equation}
        % we can confirm the leading term in the first line.
        % To check that the second line corresponds to the subleading term of the soft graviton theorem, simply notice that
        % \begin{equation}
        %     J^{\nu \rho} A_1 = - K^{\nu \rho} A_1 = (u^\nu e_1^\rho - e_1^\nu u^\rho) u e_1.
        % \end{equation}

        In the case of pure gravity and for tree amplitudes, gauge invariance also determines the next-to-next-to-leading order in the soft limit, i.e. the sub-subleading soft graviton limit.
        Naturally we may wonder if that is also the case here.
        When attempting the same scheme we run into difficulties due to the appearance of a term $\partial_\omega^2 A_{g^n w}(0; k_1 \dots k_n)$ which is absent in the pure graviton theory.
        Indeed in \eqref{eq:LON} we were able to simplify the partial derivative using the subleading soft-worldline theorem, however simplification of a second order partial derivative would necessitate a sub-subleading soft-worldline theorem which we do not have.
        This can be put in perspective with the fact that unlike the leading and subleading terms, the sub-subleading contribution is not universal: it receives loop corrections in pure gravity~\cite{Bern2014}, and is further modified by higher-order operators in an EFT treatment of gravity~\cite{Elvang2017}.

%% file: diagrams/np1-graviton-singular-a.tex
\begin{tikzpicture}[baseline={(current bounding box.center)}]
    \begin{feynman}
        
        % Worldline through blob
        \vertex (wl) at (-1.2, .15);
        \vertex (wr) at (1.2, .15);
        \draw[thick, dotted] (wl) -- (wr);
        
        % Blob
        \node[draw, circle, fill=gray!30, minimum size=1cm] (blob) {};
        
        % Graviton legs (bottom)
        \vertex (k1) at (-.7, -1.1);
        \vertex (k2) at (-.3, -1.1);
        %\node at (.15, -.9) {$. \; .$};
        \vertex (kn) at (.65, -1.1);
        \draw[thick, photon] (blob.215) -- (k1);
        \draw[thick, photon] (blob.240) -- (k2);
        \node at (.15, -.9) {$\cdots$};
        \draw[thick, photon] (blob.310) -- (kn);
        
        % Splitting vertex, placed below the worldline to avoid superposition
        \vertex (v)  at (.9, -.5);
        \vertex (ki) at (1.35, -1);
        \vertex (q)  at (1.5, -.3);
        \draw[thick, photon] (blob.335) -- (v);
        \draw[thick, photon] (v) -- (ki);
        \draw[thick, photon] (v) -- (q);
        
        % Labels
        \node at (1.55, -1.15)  {$k_i$};
        \node at (1.7, -.3) {$q$};
        
        \node at (-.75, -1.35) {$k_1$};
        \node at (-.3,  -1.35) {$k_2$};
        \node at (.25,  -1.35) {$\cdots$};
        \node at (.8,  -1.35) {$k_n$};
        
    \end{feynman}
\end{tikzpicture}

%% file: diagrams/np1-graviton-singular-b.tex
\begin{tikzpicture}[baseline={(current bounding box.center)}]
    \begin{feynman}
    
        % Worldline through blob
        \vertex (wl) at (-1.2, .15);
        \vertex (wr) at (1.6, .15);
        \vertex (u) at (0, .15);
        \vertex (v) at (1.2,.15);
        \draw[thick, dotted] (wtl) -- (wr);
        \draw[thick] (u) -- (v);
        %\draw[thick] (v) -- (wtr);

        % Grey blob
        \node[draw, circle, fill=gray!30, minimum size=1cm] (blob) {};
        
        % Graviton legs (bottom)
        \vertex (k1) at (-.7, -1.1);
        \vertex (k2) at (-.3, -1.1);
        \node at (.15, -.9) {$\cdots$};
        \vertex (kn) at (.65, -1.1);
        \draw[thick, photon] (blob.215) -- (k1);
        \draw[thick, photon] (blob.240) -- (k2);
        \draw[thick, photon] (blob.315) -- (kn);
        
        % Soft leg
        \node [circle, fill=black, inner sep=1pt] at (v) {};
        \vertex (q) at (1.4,-.65);
        \draw[thick, photon] (v) -- (q);

        % Labels
        \node at (1.4, -.9)  {$q$};
        \node at (-.75, -1.35) {$k_1$};
        \node at (-.3,  -1.35) {$k_2$};
        \node at (.25,  -1.35) {$\cdots$};
        \node at (.8,  -1.35) {$k_n$};

    \end{feynman}
\end{tikzpicture}

%% file: sections/05-recursion-algorithm.tex
\section{Recursion relations}\label{sec:recursion}

    In this section, we will finally build a set of on-shell recursion relations for WQFT amplitudes.
    Specifically, in this section we present a systematic procedure to obtain any WQFT rational amplitude, i.e. for an arbitrary number of external gravitons and worldline fluctuations.
    The leading soft-worldline theorem directly gives us the on-shell amplitude with an external worldline from the amplitude without it.
    Iterations of the formula would then give us the amplitude with any number of worldline fluctuations from the corresponding amplitude with only external gravitons.
    Hence for our purpose, a set of recursion relations for amplitudes with any number of external gravitons is sufficient to obtain any rational amplitude.

    We start with a general discussion of how and why soft theorems can help us design recursion relation in theories with bad large-$z$ behavior.
    The $n$-graviton recursion that follows is an all-line soft recursion that works in $d$-dimensions, for $n \geq 3$.
    For completeness we also give a specific recursion for the case $n=2$, but the two-graviton amplitude can very well just be taken as an input value for the general case.

    \subsection{On-shell recursion from soft theorems}\label{sec:motivation}
    
        On-shell recursion is a procedure for systematically expressing tree level amplitudes in terms of amplitudes with fewer external legs. 
        The strategy is to relate an amplitude to its singular points in the space of complex kinematics.
        From there we use the factorization properties of the amplitude to relate these special points to simpler amplitudes. 

        More explicitly, starting from a set of on-shell external momenta $p_i$, one performs a complex deformation 
        \begin{equation}\label{eq:soft_def}
            p_i \rightarrow \Tilde{p}_i(z) = p_i + z q_i.
        \end{equation}
        The shifted amplitude then becomes a function of the complex variable $z$
        \begin{equation}
            A := A(p_1,\dots,p_n) \rightarrow A(z).
        \end{equation}
        The vectors $q_i$ must be chosen such that the set of momenta $\Tilde{p}_i(z)$ satisfy the on-shell conditions at any value of~$z$.
        Different choices of $q_i$ satisfying these constraints lead to different recursion relations.
        Consequently, $A(z)$ is an on-shell and gauge invariant quantity everywhere on the complex plane.

        As a rational function of the external kinematics, $A(z)$ is meromorphic in $z$. 
        This allows us to relate its value at $z = 0$ to a contour integral via Cauchy's theorem
        \begin{equation}
            A = \frac{1}{2 \pi i} \oint_{z=0} \frac{dz}{z} A(z) = - \sum_I \Res_{z = z_I} \left[ \frac{A(z)}{z} \right] + R_\infty,
        \end{equation}
        where $z_I$'s denote all the poles of $A(z)$ and $R_\infty$ denotes the possible integral contribution when we push the contour to infinity.

        Apart  from $z=0$, the poles in the function $A(z)/z$ must correspond to a physical factorization channel with some deformed propagator going on shell.
        Unitarity implies that each of these residues can be expressed as a product of simpler on-shell amplitudes 
        \begin{equation}\label{eq:factorization}
            \Res_{z = z_I} \left[ \frac{A(z)}{z} \right] = \frac{A_L(z_I) A_R(z_I)}{P(z_I)},
        \end{equation}
        where $P$ is a polynomial in $z$ whose coefficients are functions of the (unshifted) external momenta. 
        Applying the procedure repeatedly, one can compute any on-shell amplitude starting from a small number of elementary amplitudes.
        This approach encountered many successes most commonly illustrated by the BCFW recursion for gauge theory and gravity \cite{Britto2005a, Britto2005b}.

        The main challenge in this procedure is the presence of the boundary integral $R_\infty$ and the key feature of a good complex deformation is to make $R_\infty$ vanish.
        Mathematically if our Cauchy-theorem integrand decays strictly faster than $z^{-1}$, we are guaranteed to have $R_\infty = 0$.
        % For a wide range of physically interesting theories however, poor large-$z$ behavior makes finding such a deformation difficult.
        % These include EFTs with higher-dimensional operators and in particular WQFT, which is the main focus of this article.
        Many theories however have a poor large-$z$ behavior and there are several different approaches to dealing with the residue at infinity~\cite{Jin2015,Benincasa2012,Feng2010}.
        Here we take inspiration from~\cite{Luo2016, Cheung2016}, which exploits the soft behavior of the theory to improve the behavior of the amplitudes at infinity.

        The general idea is straightforward: assume that for some particular value $z_0$ of the complex parameter, one of the deformed momenta vanishes $p_i(z_0)=0$ and consider the following integral
        \begin{equation}
            A = \frac{1}{2 i \pi} \oint_{z=0} \frac{dz}{z} \frac{A(z)}{(1-z/z_0)}.
        \end{equation}
        Clearly the value of the residue of the pole in $z=0$ is left unchanged.
        Expressing the new contour at infinity $R'_\infty$ as the sum of all the residues we get a new expression for $A$
        \begin{equation}
            A = - \sum_I \Res_{z = z_I} \left[ \frac{A(z)}{z(1-z/z_0)} \right] - \Res_{z = z_0} \left[ \frac{A(z)}{z(1-z/z_0)} \right] + R'_\infty.
        \end{equation}
        Notice that as $\vert z \vert \rightarrow \infty$, the new integrand has an additional power of $z$ in the denominator.
        This improvement of the large-$z$ behavior of our integrand comes at the price of introducing a new pole in $z_0$. 
        But because it corresponds to the vanishing of an external momentum, its residue is precisely the information provided by a soft theorem.
        More generally, higher powers of $(1-z/z_0)$ in the denominator would further improve the large-$z$ behavior (and eventually force $R'_\infty = 0$), at the cost of introducing subleading soft poles at $z_0$.
        These correspond to subleading soft theorems.
        Worldline amplitudes with $m$ external worldline fluctuations generically behave as $z^m$ at large $z$, making some infrared input essential to the construction of a recursion formula.
    
    \subsection{\texorpdfstring{$n \geq 3$}{n >= 3} graviton amplitude}\label{sec:general-recursion}
        
        We now formulate our recursion formula for a general $n \geq 3$ graviton rational amplitude $A_{g^n}^{\text{WQFT}}(k_1\dots k_n)$ in the worldline theory.
        Our recursion starts with a soft deformation on all external momenta
        \begin{equation}\label{eq:shift}
            k_i \rightarrow \Tilde{k}_i(z) = k_i (1-z/z_i),
        \end{equation}
        and we define the shifted amplitude
        \begin{equation}
            A_n(z) := A_{g^n}^{\text{WQFT}}\big(\Tilde{k}_1(z)\dots \Tilde{k}_n(z)\big).
        \end{equation}
        In a QFT with full Poincaré symmetry, the requirement that this quantity be on shell for any complex $z$ would require global momentum conservation $\sum_i k_i^\mu/z_i = 0$.
        With general kinematics, this equation has non-trivial solutions for $z_i$ only when $n \geq d+2$.
        Here however, translation symmetry is broken by the background value of the worldline.
        Hence global momentum conservation is relaxed to global energy conservation in the $u$ frame.
        For our shift this translates to a single constraint $\sum_{i = 1}^n (u \! \cdot \!k_i)/z_i = 0$, which has non-trivial solutions\footnote{The case $n = 2$ forces $z_1 = z_2$ and since amplitudes are homogeneous in momenta, the shifted amplitude would be $A(z) = A(0)(1-z/z_{1,2})$. As this deformation does not explore any non-trivial kinematic sector of the amplitude, we cannot use it for a recursion. Consequently we require that $n \geq 3$, so we are guaranteed a non-trivial solution.} 
            for $z_i$ as long as $n \geq 3$.

        Using the soft-graviton theorem derived in Sec.~\ref{sec:soft-graviton}, we can access the first two terms in the Laurent series expansion of $A_n(z)$ around a soft pole.
        Indeed, around the point $z = z_i$, the shifted momentum $\tilde{k}_i(z)$ becomes soft.
        Therefore expanding the soft-graviton theorem~\eqref{eq:soft-graviton-theorem} to the corresponding order in $(z-z_i)$ directly yields the residue and finite part
        \begin{equation}\label{eq:laurent-series}
            A_n(z) = \frac{\mathcal{S}^{(\minus 1)}_i}{z-z_i} + \mathcal{S}^{(0)}_i + \mathcal{O}(z-z_i).
        \end{equation}
        These coefficients are given explicitly in term of the graviton amplitude with one fewer leg
        \begin{equation}\label{eq:laurentCoef}
        \begin{aligned}
            \mathcal{S}^{(\minus 1)}_i &= \lim_{z \to z_i}(z-z_i)\left(S^{(\minus 1)}_i A_{n-1}\right)(z), \\
            \mathcal{S}^{(0)}_i &= \lim_{z \to z_i} \left( \left(\big[S^{(\minus 1)}_i + S^{(0)}_i\big] A_{n-1}\right)(z) - \frac{\mathcal{S}^{(\minus 1)}_i}{z-z_i} \right).
                \end{aligned}
        \end{equation}
        An expression like $\left(\big[S^{(\minus 1)}_i + S^{(0)}_i\big] A_{n-1}\right)(z)$ means we first apply the leading and subleading soft operators to the ($n{-}1$)-graviton amplitude (with a possible relabeling of the indices so that the $i$th momentum is soft) and then perform the complex deformation~\eqref{eq:shift} on the result.
        
        As explained above, power counting and a direct analysis of the Feynman rules shows that rational amplitudes in WQFT behave like $A_{g^n w^m} \sim z^m $ at large-$z$.
        So following the discussion in Sec.~\ref{sec:motivation}, for an all-graviton amplitude ($m=0$) we only need a single soft factor to remove any potential boundary contribution when the contour is pushed to infinity.
    
        With that in mind we express our unshifted amplitude as the following contour integral
        \begin{equation}\label{eq:wqft-cauchy}
            A_n(0) = \frac{1}{2 \pi i} \oint_{z=0} \frac{dz}{z} \frac{A_n(z)}{(1-z/z_1)}.
        \end{equation}
        Of course the choice of $z_1$ is arbitrary and one could have chosen any other $i \neq 1$ for the additional soft factor of the integrand.
        Before we evaluate this integral, it is useful to classify the residues into three categories, relative to the singularity that produces them.
        
        \paragraph{Soft poles.} 
        By construction, the $i$th leg becomes soft exactly as $z = z_i$, where we have $\tilde{k}_i(z_i)=0$.
        The amplitude has a simple pole there, whose residue is fixed by the soft graviton theorem through the Laurent coefficients $\mathcal{S}^{(\minus 1)}_i$ and $\mathcal{S}^{(0)}_i$ in~\eqref{eq:laurent-series}.
        There are $n$ such poles, one per external leg. 
        The pole at $z_1$ is special: it coincides with the extra factor of $(1-z/z_1)^{-1}$, so the two combine into a double pole whose residue involves both $\mathcal{S}^{(\minus 1)}_i$ and $\mathcal{S}^{(0)}_i$.

        \paragraph{Graviton factorization poles.}
        Internal graviton propagators $k_I^{2}$, labeled by a non-repeating multi-index $I$ that will be made more explicit later, can go on shell under the deformation.
        As $k_I^{2}$ becomes quadratic in $z$ once shifted, each such propagator contributes two simple poles $z_I^{\pm}$, given in \eqref{eq:factorization_poles}.

        \paragraph{Worldline factorization poles.}
        Internal worldline propagators $(u \! \cdot \! k_J)^{2}$ also labeled by a multi-index $J$ can equally go on-shell.
        Since $u \! \cdot \! \tilde{k}_J(z)$ is linear in $z$, the shifted propagator produces a double pole at $z_J$.
        The residues involve the $z$-derivative of the sub-amplitudes on either side, which remain well defined on-shell quantities as we will explain below \eqref{eq:recursion}.

        Poles adjacent to an external leg are already captured by the soft residues and must be excluded from both factorization sums, to avoid double-counting.
        
        Collecting all three contributions, Cauchy's theorem applied to the contour integral~\eqref{eq:wqft-cauchy} gives
        % \begin{multline}\label{eq:recursion}
        %     A_n(0) \; =
        %     \quad  \sum_I \frac{1}{k_I^2} \frac{A_\text{W}(z_I^+)A_\text{EH}(z_I^+)}{(1-z_I^+/z_I^-)(1-z_I^+/z_1)} \quad + \quad(z_I^+ \leftrightarrow z_I^-) \\
        %     - \frac{1}{m} \sum_J \frac{1}{(u k_J)^2} \left\{ \frac{(1-2 z_J/z_1)A_\text{L}(z_J) A_\text{R}(z_J)}{z_J^2(1-z_J/z_1)^2} - \frac{A_\text{L}'(z_J) A_\text{R}(z_J) + A_\text{L}(z_J) A_\text{R}'(z_J)}{z_J(1-z_J/z_1)} \right\} \\
        %     -\frac{\mathcal{S}^{(\minus 1)}_1}{z_1} + \mathcal{S}^{(0)}_1 - \sum_{i=2}^n \frac{\mathcal{S}_i^{(\minus 1)}}{z_i(1-z_i/z_1)},
        % \end{multline}
        \begin{equation}\label{eq:recursion}
        {\setlength{\fboxsep}{8pt}
        \boxed{
        \begin{aligned}
            A_n(0) \; &=
            \quad  \sum_I \frac{1}{k_I^2} \frac{A_\text{W}(z_I^+)A_\text{EH}(z_I^+)}{(1-z_I^+/z_I^-)(1-z_I^+/z_1)} \quad + \quad(z_I^+ \leftrightarrow z_I^-) \\
            - \frac{1}{m}& \sum_J \frac{1}{(u \! \cdot \! k_J)^2} \left\{ \frac{(1-2 z_J/z_1)A_\text{L}(z_J) A_\text{R}(z_J)}{z_J^2(1-z_J/z_1)^2} - \frac{A_\text{L}'(z_J) A_\text{R}(z_J) + A_\text{L}(z_J) A_\text{R}'(z_J)}{z_J(1-z_J/z_1)} \right\} \\
            & \qquad\qquad\qquad\qquad\qquad\qquad\qquad\qquad -\frac{\mathcal{S}^{(\minus 1)}_1}{z_1} + \mathcal{S}^{(0)}_1 - \sum_{i=2}^n \frac{\mathcal{S}_i^{(\minus 1)}}{z_i(1-z_i/z_1)},
        \end{aligned}
        }}
        \end{equation}
        where the sums over $I$ and $J$ run over a subset of all the factorization poles as we will explain shortly.
        Summation over the physical polarizations is implicit; for gravitons it is done the usual way by replacing the internal polarization $e_{\mu \nu} e_{\rho \sigma} \to P_{\mu \nu; \rho \sigma}$. 
        For the worldline we do $\zeta_\mu \zeta_\nu \to \eta_{\mu \nu}$.
        This equation represents the main result of this article and we now explain the meaning of each line in more details.
        
        In last line we recognize the contributions from the soft residues that naturally involves the coefficients of the soft Laurent series~\eqref{eq:laurent-series}.
        We see that the residue in $z_1$ has a special value reflecting our choice of denominator in~\eqref{eq:wqft-cauchy}, and involves the subleading Laurent coefficient $\mathcal{S}^{(0)}_1$.

\begin{figure}[t]
        \centering
        \begin{subfigure}[]{0.3\textwidth}
            \centering
            \input{diagrams/graviton-factorization}
            \caption{}
            \label{fig:factorization_graviton}
        \end{subfigure}
        \begin{subfigure}[]{0.3\textwidth}
            \centering
            \input{diagrams/z-factorization}
            \caption{}
            \label{fig:factorization_worldline}
        \end{subfigure}
        \caption{Factorization channels of the $n$-graviton amplitude, corresponding to graviton and worldline fluctuation internal propagators going on-shell.}
        \end{figure}
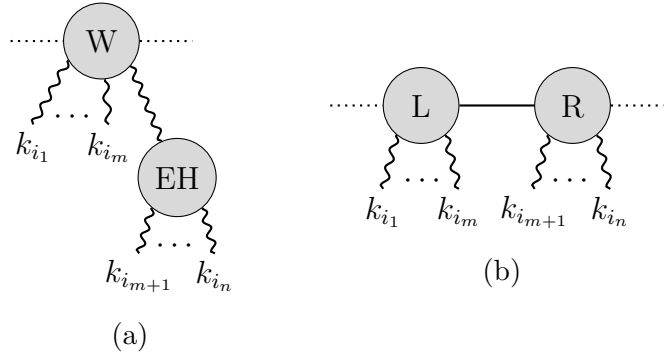

        The first line in the formula corresponds to factorization poles arising from a shifted graviton propagator going on shell, as represented in Fig.~\ref{fig:factorization_graviton}.
        Any such propagator is uniquely labeled by an ordered, non-repeating multi-index $I = (i_1, i_2 \dots i_m)$ in the set $\{1,2\dots n \}$. 
        We write $\vert I \vert := m$ and define
        \begin{equation}
            k_I := k_{i_1} + k_{i_2} + \dots + k_{i_m}.
        \end{equation}
        In a $n$-point amplitude there are $2^n-(n{+}1)$ distinct graviton propagators.
        Notice that when a propagator (graviton or worldline fluctuation) is adjacent to an external leg, the corresponding residue is already present as a contribution to one of the soft residues.
        To avoid overcounting, these should not be included in the factorization sums.
        In particular for graviton propagators, the sum should only include factorization poles for which $\vert I \vert \geq 3$.
        When shifted, a graviton propagator $k_I^2$ becomes a second order polynomial in $z$, whose roots
        \begin{equation}\label{eq:factorization_poles}
            z_I^\pm = \frac{(\sum_{i \in I} k_i/z_i) \! \cdot \! k_I \pm \sqrt{[(\sum_{i \in I} k_i/z_i) \! \cdot \! k_I]^2 - k_I^2 (\sum_{i \in I} k_i/z_i)^2} }{(\sum_{i \in I} k_i/z_i)^2},
        \end{equation}
        give the location of the corresponding poles in the complex plane.
        The factorization residue of a graviton propagator involves both a pure gravity amplitude $A_{\text{EH}}$ and a WQFT amplitude $A_{\text{W}}$ with $\vert I \vert +1$ and $n - \vert I \vert +1$ external graviton respectively.
        For illustration, the three-point pure gravity amplitude is
        \begin{equation}\label{eq:3-EH}
            A_3^{\text{EH}} = - \kappa [(e_1 \! \cdot \! e_2)(e_3 \! \cdot \! p_1) + (e_2 \! \cdot \! e_3)(e_1 \! \cdot \! p_2) - (e_1 \! \cdot \! e_3)(e_2 \! \cdot \! p_1)]^2.
        \end{equation}
        
        Finally the second line in \eqref{eq:recursion} is the sum over factorization residues corresponding to a worldline propagator
        \begin{equation}
            (u \! \cdot \! k_J)^2 := (u \! \cdot \! k_{j_1} + u \! \cdot \! k_{j_2} + \dots + u \! \cdot \! k_{j_m})^2
        \end{equation}
        going on-shell, as depicted in Fig.~\ref{fig:factorization_worldline}.
        Again to avoid overcounting contributions that are already present in the soft residues, we should exclude the case $\vert J \vert = 1$ from the sum.
        Because of energy conservation, any such propagator will be equal to its complementary propagator (e.g. for $n=4$ we have $(u \! \cdot \! k_1 + u \! \cdot \! k_2)^2 = (u \! \cdot \! k_3 + u \! \cdot \! k_4)^2$).
        Hence for odd $n$ we consider only those poles with $2 \leq \vert J \vert < n/2$ --- note the absence of this contribution for $n=3$.
        When $n$ is even on the other hand, we consider poles with $2 \leq \vert J \vert < n/2$ and with $\vert J \vert=n/2$ that include $k_1$.
        The expression of the factorization residue is more involved than the simple form advertised in Eq.~\eqref{eq:factorization}. 
        This is because shifting a worldline propagator introduces a double pole, located at
        \begin{equation}
            z_J = \frac{u \! \cdot \! k_J}{\sum_{j \in J}u \! \cdot \! k_j /z_j}.
        \end{equation}

        Naively the derivatives $A_{L,R}'(z_I)$ seem to be off-shell quantities.
        Indeed $\omega(z) = u \! \cdot \!k_J(z)$ vanishes at $z=z_J$ but not in general.
        Thus taking the complex derivative, the chain rule would involve a suspicious partial derivative $\partial_\omega$.
        However our subleading soft-worldline theorem \eqref{eq:subleading-soft-z} precisely states that this quantity is well defined on-shell.
        By first applying the chain rule, then using both our soft-$z$ theorems we have
        \begin{equation}
            A_{g^m w}'(z_J) = i \bigg[\sum_{j \in J} \bigg( \frac{u \! \cdot \! k_j}{z_j} \zeta^\mu \frac{\partial}{\partial u^\mu} - \frac{\zeta \! \cdot \! k_j}{z_j} - \zeta \! \cdot \! \tilde{k}_J(z_J) \frac{k_j^\mu}{z_j} \frac{\partial}{\partial k_j^\mu} \bigg)\bigg] A_m(z_J),
        \end{equation}
        where $\tilde{k}_J(z_J) = \sum_{j \in J}\tilde{k}_j(z_J)$.
        This makes apparent the fact that $A_{L,R}'(z_I)$ is just an on-shell operator\footnote{in the sense that applying this operator to an expression that is zero on-shell will yield zero. From a geometric perspective, this operator is tangent to the surface defined by the on-shell conditions.} applied to an all-graviton WQFT amplitude with $m < n$.
        In that regard it does not cause any problem and fits nicely in the recursion.

        The recursion~\eqref{eq:recursion} constructs the amplitudes of the action where all the Wilson coefficients discussed below Eq.~\eqref{eq:wl-action} are set to zero.
        This is a consequence of our assumption that every all-graviton amplitude in the theory scales as $\mathcal{O}(z^0)$, which holds for this minimal theory but fails once higher-dimensional operators are included.

    \subsection{The case \texorpdfstring{$n = 2$}{n = 2}}
    
        We have argued that the recursion we have outlined does not work for $n = 2$.
        This is not really an issue as we can just take the two-point amplitude as an input of the recursion.
        However, as promised, we now give a deformation that works specifically for $n=2$.
        We define the complex kinematic shift
        \begin{equation}\label{eq:n=2_deformation}
        \begin{aligned}
            k_1 \rightarrow k_1 - \frac{z(u \! \cdot \! k_1)}{z_0 (u \! \cdot \! e_1)} e_1, \\
            k_2 \rightarrow k_2 (1-z/z_0),
        \end{aligned}
        \end{equation}
        where $e_1$ is the complex null polarization of $k_1$.
        The key insight here is that the first leg is not simply rescaled, but is shifted in the direction of its polarization so that we avoid the issue mentioned before.
        A direct computation confirms that the deformed momenta satisfy all on-shell conditions and energy is globally conserved.
        
        We have the same large-$z$ behavior as before $A_{2}(z) \sim z^0$, and in order to avoid a boundary contribution we write Cauchy's theorem as 
        \begin{equation}
            A_2(0) = \frac{1}{2 \pi i} \oint \frac{A_2(z)}{z(1-z/z_0)} dz.
        \end{equation}
        This complex shift only probes the soft region of the second leg, and we only have a single Laurent series from which we deduce the soft residue
        \begin{equation}
            A_2(z) = \frac{\mathcal{S}^{(\minus 1)}}{z-z_0} + \mathcal{S}^{(0)} + \mathcal{O}(z-z_0) \quad \longrightarrow \quad \Res_{z = z_0} \left[ \frac{A_2(z)}{z(1-z/z_0)} \right] = \frac{\mathcal{S}^{(\minus 1)}}{z_0} - \mathcal{S}^{(0)}.
        \end{equation}
        There are two possible propagators in the amplitude. 
        The factorization pole arising from deforming $(u \! \cdot \! k_1)^2 = (u \! \cdot \! k_2)^2$ is already included in the soft residue.
        The other propagator deforms to
        \begin{equation}
            2(k_1 \! \cdot \! k_2)(1-z/z_0)(1-z/z_1) \quad \text{with} \quad z_1 = z_0 \frac{(u \! \cdot \! e_1)(k_1 \! \cdot \! k_2)}{(u \! \cdot \! k_1)(e_1 \! \cdot \! k_2)}.
        \end{equation}
        From this we can compute the last residue to find the final expression of our amplitude
        \begin{equation}\label{eq:n=2}
            A_{2} = - \frac{\mathcal{S}^{(\minus 1)}}{z_0} + \mathcal{S}^{(0)} + \frac{A_{1}(z_1) A_{3}^{\text{EH}}(z_1)}{2(k_1 \! \cdot \! k_2) (1-z_1/z_0)^2},
        \end{equation}
        where $A_{1}$ is the one-graviton vertex of WQFT and $A_{3}^{\text{EH}}$ is the three-graviton vertex of pure gravity.

%% file: diagrams/graviton-factorization.tex
\begin{tikzpicture}[baseline={(current bounding box.center)}]
    \begin{feynman}

        % Grey blob
        \node[draw, circle, fill=gray!30, minimum size=1cm] (blob1) {W};
        \node[draw, circle, fill=gray!30, minimum size=1cm] (blob2) at (1,-1.8) {EH};

        % Graviton legs
        \vertex (u1) at (-1.2,0);
        \vertex (u2) at (1.2,0);

        \vertex (k1) at (-.9,-1.1);
        \node at (-.35,-1) {$\cdots$};
        \vertex (k2) at (.1,-1.1);

        \vertex (k3) at (.5,-2.8);
        \node at (1,-2.7) {$\cdots$};
        \vertex (k4) at (1.5,-2.8);

        \draw[thick, dotted] (u1) -- (blob1);
        \draw[thick, dotted] (blob1) -- (u2);

        \draw[thick, photon] (blob1.210) -- (k1);
        \draw[thick, photon] (blob1) -- (k2);
                        
        \draw[thick, photon] (blob1.320) -- (blob2);
                        
        \draw[thick, photon] (blob2.230) -- (k3);
        \draw[thick, photon] (blob2.310) -- (k4);

        % Labels
        \node at (-.9,-1.4) {$k_{i_1}$};
        \node at (.1,-1.4) {$k_{i_m}$};
        \node at (.5,-3.1) {$k_{i_{m+1}}$};
        \node at (1.5,-3.1) {$k_{i_{n}}$};

    \end{feynman}
\end{tikzpicture}

%% file: diagrams/z-factorization.tex
\begin{tikzpicture}[baseline={(current bounding box.center)}]
    \begin{feynman}

        % Grey blob
        \node[draw, circle, fill=gray!30, minimum size=1cm] (blob1) {L};
        \node[draw, circle, fill=gray!30, minimum size=1cm] (blob2) at (2,0) {R};

        % Graviton legs
        \vertex (u1) at (-1.2,0);
        \vertex (u2) at (3.2,0);
                        
        \vertex (k1) at (-.5,-1.1);
        \node at (.05,-1) {$\cdots$};
        \vertex (k2) at (.5,-1.1);
                        
        \vertex (k3) at (1.5,-1.1);
        \node at (2,-1) {$\cdots$};
        \vertex (k4) at (2.5,-1.1);
                        
        \draw[thick, dotted] (u1) -- (blob1);

        \draw[thick, photon] (blob1.230) -- (k1);
        \draw[thick, photon] (blob1.310) -- (k2);
                        
        \draw[thick] (blob1) -- (blob2);
                        
        \draw[thick, photon] (blob2.230) -- (k3);
        \draw[thick, photon] (blob2.310) -- (k4);
                        
        \draw[thick, dotted] (blob2) -- (u2);

        % Labels
        \node at (-.5,-1.4) {$k_{i_1}$};
        \node at (.5,-1.4) {$k_{i_m}$};
        \node at (1.5,-1.4) {$k_{i_{m+1}}$};
        \node at (2.5,-1.4) {$k_{i_{n}}$};
                        
    \end{feynman}
\end{tikzpicture}

%% file: sections/06-results.tex
\section{Worked examples}\label{sec:results}

    Let us now go through some concrete application of these formulas.
    We begin with the $n=2$ recursion as the expressions are more manageable.
    They quickly become very long for $n=3$, so we will only describe a few intermediate results with some complementary explanations.

    \subsection{Two-graviton}

        In this case, the recursion formula contains three distinct terms. 
        Equation~\eqref{eq:laurentCoef} describes how to evaluate the first two, which correspond to the residue of the soft pole.
        In practice for computing $\mathcal{S}^{(\minus 1)}$, take the leading term of the soft graviton theorem (the first line in~\eqref{eq:A2_soft_expansion}), apply the complex deformation of Eq.~\eqref{eq:n=2_deformation} and compute the obtained leading coefficient of the Laurent series
        % \begin{equation}
        %     \mathcal{S}^{(-1)} = z_{0}\frac{ \left( u e_{2} \left( u p_{1} ((e_{1} e_{2})(u p_{2}) - (e_{1} p_{2})(u e_{2})) + u e_{1} ((p_{1} p_{2})(u e_{2}) - (e_{2} p_{1})(u p_{2})) \right) \right)^{2}}{4 (u p_{1})^{2} \left( (e_{2} p_{1})(u p_{2}) - (p_{1} p_{2})(u e_{2}) \right)}.
        % \end{equation}
        \begin{equation}
            \mathcal{S}^{(\minus 1)} = z_0 \frac{u \! \cdot \! e_1 \mathcal{K}_{12}^2}{4(u \! \cdot \! k_1)^2 [(k_{1} \! \cdot \! k_{2})(u \! \cdot \! e_{1}) - (e_{1} \! \cdot \! k_{2})(u \! \cdot \! k_{1})]},
        \end{equation}
        with $\mathcal{K}_{12}$ defined in \eqref{eq:k_factor}.
        Note that $z_0$ will cancel out when we plug this expression into the recursion.
        Using the same method we compute the subleading Laurent coefficient
        % \begin{multline}
        %     \mathcal{S}^{(0)} = \frac{u e_{2}}{4} \bigg(  2 u e_{1} \left( \frac{(e_{1} e_{2}) (u p_{2})}{u p_{1}} + \frac{((e_{1} p_{2})(e_{2} p_{1}) + (e_{1} e_{2})(p_{1} p_{2})) (u e_{2}) - 2 (e_{1} e_{2})(e_{2} p_{1})(u p_{2})}{(p_{1} p_{2})(u e_{2}) - (e_{2} p_{1})(u p_{2})} \right) \\
        %     -\frac{(e_{2} p_{1}) (u e_{1})^{2} (2 (u p_{1}) + (u p_{2}))}{(u p_{1})^{2}}
        %     - \frac{1}{((e_{2} p_{1})(u p_{2}) - (p_{1} p_{2})(u e_{2}))^{2}}
        %     ((e_{1} e_{2})(u p_{2}) - (e_{1} p_{2})(u e_{2}))  \\
        %     \times( 2 (e_{1} e_{2})(u p_{1}) ((e_{2} p_{1})(u p_{2}) - (p_{1} p_{2})(u e_{2})) + u p_{2} ((e_{1} e_{2})(e_{2} p_{1})(u p_{2}) + ((e_{1} p_{2})(e_{2} p_{1}) - 2 (e_{1} e_{2})(p_{1} p_{2})) u e_{2}) ) \bigg).
        % \end{multline}
        \begin{equation}
            \mathcal{S}^{(0)} = - \frac{(e_1 \! \cdot \! k_2)(u \! \cdot \! e_1) \mathcal{K}_{12}^2}{4u \! \cdot \! k_1 [(k_{1} \! \cdot \! k_{2})(u \! \cdot \! e_{1}) - (e_{1} \! \cdot \! k_{2})(u \! \cdot \! k_{1})]^2}.
        \end{equation}

        The remaining term in the recursion~\eqref{eq:n=2} is the factorization residue. 
        For computing the numerator we first evaluate the product of the unshifted amplitudes~\eqref{eq:g1-wqft} and~\eqref{eq:3-EH}. 
        After a convenient relabeling of the indices, we perform the implicit summation over the physical polarizations of the on-shell, internal graviton line
        \begin{equation}
            \sum_{\text{pol.}}A_{1}A_{3}^{\text{EH}} = \input{diagrams/graviton_cut} = -\frac{1}{2}[(e_1 \! \cdot \! e_2)(u \! \cdot \! k_1) + (e_1 \! \cdot \! k_2)(u \! \cdot \! e_2) - (e_2 \! \cdot \! k_1)(u \! \cdot \! e_1)]^2,
        \end{equation}
        and only then do we apply the complex deformation with $z = z_1$ to find
        \begin{equation}
            A_{1}(z_1)A_{3}^{\text{EH}}(z_1) = - \frac{(e_1 \! \cdot \! k_2)^2\mathcal{K}_{12}^2}{4k_1 \! \cdot \! k_2[(k_{1} \! \cdot \! k_{2})(u \! \cdot \! e_{1}) - (e_{1} \! \cdot \! k_{2})(u \! \cdot \! k_{1})]^2}.
        \end{equation}
        Combining the three expressions we obtain the two-graviton amplitude presented in Eq.~\eqref{eq:2gAmp}.
        Importantly the final result is independent of the choice of $z_0$, the location of the soft pole in the complex plane.

        We will see in the next section that expressions become very long quickly as $n$ increases.
        A practical way to write compact expressions is then to specialize to $d = 4$ and use the spinor-helicity variables.
        Appendix~\ref{app:spinor-helicity} explains how to adapt the formalism to worldline kinematics.
        The two-graviton, WQFT amplitude written in this formalism becomes
        \begin{equation}\label{eq:2point}
            A(1^+ 2^+) = \frac{[12]^4}{(k_1 \! \cdot \! k_2)(u \! \cdot \! k_1) (u \! \cdot \! k_2)}, 
            \qquad 
            A(1^- 2^+) = \frac{\langle 1 \vert u \vert2]^4}{(k_1 \! \cdot \! k_2)(u \! \cdot \! k_1) (u \! \cdot \! k_2)},
        \end{equation}
        Amplitudes with opposite polarities are obtained by exchanging square and angle brackets.
        These expressions were simplified to be free of any spurious poles and they manifestly have the expected permutation symmetries.

    \subsection{Three-graviton}

        Specializing the general recursion formula to $n=3$ gives
        \begin{multline}\label{eq:n=3}
            A_3(0) = \frac{1}{(k_1 + k_2 + k_3)^2} \left( \frac{A_1(z_+) A^{\text{EH}}_4(z_+)}{(1-z_+/z_-)(1-z_+/z_1)} + \frac{A_1(z_-) A^{\text{EH}}_4(z_-)}{(1-z_-/z_+)(1-z_-/z_1)} \right) \\
            -\frac{\mathcal{S}^{(\text{-}1)}_1}{z_1} + \mathcal{S}^{(0)}_1 - \frac{\mathcal{S}_2^{(\text{-}1)}}{z_2(1-z_2/z_1)} - \frac{\mathcal{S}_3^{(\text{-}1)}}{z_3(1-z_3/z_1)}.
        \end{multline}
        Notice the absence of the worldline factorization residue in the second line of~\eqref{eq:recursion}.
        Indeed, as explained earlier, this term starts contributing only at $n=4$.
        
        The leading term in the soft Laurent expansion around $z_1$ is given explicitly as
        \begin{equation}
            \mathcal{S}_1^{(\minus 1)} = - \frac{\kappa}{2} z_1 \Bigg\{ \sum_{i = 2,3} \frac{1-z_1/z_i}{k_1 \! \cdot \! k_i}\left(e_1 \! \cdot \! k_i - \frac{k_1 \! \cdot \! k_i}{u \! \cdot \! k_1} u \! \cdot \! e_1 \right)^2 \Bigg\} A_2\left(\tilde{k}_2(z_1), \tilde{k}_3(z_1) \right),
        \end{equation}
        while $\mathcal{S}_2^{(\minus 1)}$ and $\mathcal{S}_3^{(\minus 1)}$ can be obtained by relabeling the indices.
        The full expression for the subleading term $\mathcal{S}_1^{(0)}$ is significantly more involved but it can be obtained directly from Eq.~\eqref{eq:laurentCoef}.
        
        Now considering the factorization residue, we need to evaluate the following cut, again with the undeformed kinematics
        \begin{multline}
            A_{1}A_{4}^{\text{EH}} = \input{diagrams/graviton_cut_2} =-\frac{1}{4 (k_{1} \! \cdot \! k_{2})(k_{1} \! \cdot \! k_{3})\left( k_{1} \! \cdot \! k_{2} + k_{1} \! \cdot \! k_{3}\right)} \Big(
            (e_{1} \! \cdot \! e_{2}) (k_{1} \! \cdot \! k_{3})^{2} (u \! \cdot \! e_{3}) 
            \\[-.4em]
            + k_{1} \! \cdot \! k_{3} \big((e_{3} \! \cdot \! k_{1}) (e_{2} \! \cdot \! k_{1}) (u \! \cdot \! e_{1}) -(e_{1} \! \cdot \! k_{3}) (e_{2} \! \cdot \! k_{1}) (u \! \cdot \! e_{3}) + (e_{3} \! \cdot \! k_{2}) (e_{2} \! \cdot \! k_{1}) (u \! \cdot \! e_{1})  
            \\ 
            + (e_{1} \! \cdot \! k_{2}) (e_{2} \! \cdot \!  k_{3}) (u \! \cdot \! e_{3}) - (e_{1} \! \cdot \! k_{2}) (e_{3} \! \cdot \! k_{1}) (u \! \cdot \! e_{2}) - (e_{1} \! \cdot \! k_{2}) (e_{3} \! \cdot \! k_{2}) (u \! \cdot \! e_{2})
            \\
            + u \! \cdot \! k_{2} \big[ (e_{2} \! \cdot \! e_{3}) (e_{1} \! \cdot \! k_{2}) - (e_{1} \! \cdot \! e_{3}) (e_{2} \! \cdot \! k_{1}) + (e_{1} \! \cdot \! e_{2}) (e_{3} \! \cdot \! k_{1}) \big]
            \\
            + u \! \cdot \! k_{1} \big[(e_{2} \! \cdot \! e_{3}) (e_{1} \! \cdot \! k_{2}) - (e_{1} \! \cdot \! e_{3}) (e_{2} \! \cdot \! k_{1}) - (e_{1} \! \cdot \! e_{2}) (e_{3} \! \cdot \! k_{2}) \big]  
            \\
            + k_{1} \! \cdot \! k_{2} \big[(e_{1} \! \cdot \! e_{2}) (u \! \cdot \! e_{3}) + (e_{1} \! \cdot \! e_{3}) (u \! \cdot \! e_{2}) - (e_{2} \! \cdot \! e_{3}) (u \! \cdot \! e_{1}) \big] \big) 
            \\
            - k_{1} \! \cdot \! k_{2} \big( (e_{1} \! \cdot \! e_{3}) (e_{2} \! \cdot \! k_{1}) (u \! \cdot \! k_{1}) + (e_{1} \! \cdot \! e_{3}) (e_{2} \! \cdot \! k_{3}) (u \! \cdot \! k_{1}) + e_{1} \! \cdot \! k_{3} ( e_{2} \! \cdot \! k_{1} + e_{2} \! \cdot \! k_{3} ) u \! \cdot \! e_{3}
            \\
            - e_{3} \! \cdot \! k_{1}( e_{2} \! \cdot \! k_{1} + e_{2} \! \cdot \! k_{3} ) u \! \cdot \! e_{1}
            + u \! \cdot \! k_{2} \big[(e_{2} \! \cdot \! e_{3}) (e_{1} \! \cdot \! k_{3}) + (e_{1} \! \cdot \! e_{3}) (e_{2} \! \cdot \! k_{1}) - (e_{1} \! \cdot \! e_{2}) (e_{3} \! \cdot \! k_{1})\big] 
            \\[-.2em]
            + (e_{1} \! \cdot \! k_{2}) (e_{3} \! \cdot \! k_{1}) (u \! \cdot \! e_{2}) - u \! \cdot \! e_{2}\big[ (e_{1} \! \cdot \! k_{3}) (e_{3} \! \cdot \! k_{2}) + (e_{1} \! \cdot \! e_{3}) (k_{1} \! \cdot \! k_{2}) \big]  \big)
            \Big)^{2}.
        \end{multline}
        From there the two numerators in~\eqref{eq:n=3} are obtained by applying the complex deformation and setting $z = z_\pm$.
        Equation~\eqref{eq:factorization_poles} introduces spurious square roots into the recursion output, which are difficult to simplify and significantly lengthen the final expressions.
        However numerical verification shows that our results coincide with the amplitudes computed through other methods~\cite{He2025}.

        Expressed in spinor-helicity variables the all-plus amplitude is
        \begin{multline}\label{eq:3point}
            A(1^+2^+3^+) = - \frac{[12]^2 [13]^2 [23]^2(9(u \! \cdot \! k_1)((u \! \cdot \! k_2)^2 + (u \! \cdot \! k_3)^2) + 2((u \! \cdot \! k_2)^3 + (u \! \cdot \! k_3)^3))}{6 (k_1 \! \cdot \! k_2)(k_1 \! \cdot \! k_3)(u \! \cdot \! k_1)(u \! \cdot \! k_2)^2(u \! \cdot \! k_3)^2}  \\
            - \frac{[12]^2[13]^3[23] \langle 1 \vert u \vert 2 ](3(u \! \cdot \! k_1)^3 - 6 (u \! \cdot \! k_2)^2(u \! \cdot \! k_1) - 4 (u \! \cdot \! k_2)^3)}{3 (k_1 \! \cdot \! k_2)(k_1 \! \cdot \! k_3)(u \! \cdot \! k_1)^2(u \! \cdot \! k_2)^2(u \! \cdot \! k_3)^2} \\
            - \frac{[12]^3[13]^3\langle 1 \vert u \vert 2 ]\langle 1 \vert u \vert 3 ]}{(k_1 \! \cdot \! k_2)(k_1 \! \cdot \! k_3)(u \! \cdot \! k_2)^2(u \! \cdot \! k_3)^2} \quad + \text{ perm.},
        \end{multline}
        where we sum over all permutations of the 3 graviton indices.
        Again the all-minus quantity is obtained by exchanging square and angle brackets.
        Although it is present in the $d$-dimensional expression, the graviton propagator $(k_1 + k_2 + k_3)^2$ is absent from any denominator of the all-plus spinor helicity expression.
        This can be understood by noting that the residue of that propagator in the spinor helicity amplitude must be a linear combination of $A_{\text{EH}}(1^+2^+3^+4^+)$ and $A_{\text{EH}}(1^+2^+3^+4^-)$, which are both known to vanish.
        The maximum helicity violation amplitudes (e.g. $(1^+2^+3^-)$) can be obtained from the $d$-dimensional amplitude, but the expression is too long to be displayed here even in spinor-helicity.

        We have mentioned that the second line of Eq.~\eqref{eq:recursion}, corresponding to a shifted worldline propagator going on-shell gave no contribution at $n=3$.
        As a sanity check we have verified that the soft residue in $z_{2}$ (equivalently $z_3$) is indeed equal to the sum over the factorization residues located at $z_2$.
        Since this sum involves a worldline factorization residue as $u \! \cdot \! k_2(z_2) = 0$, we were able to confirm that we have the correct expression.

%% file: diagrams/graviton_cut.tex
\begin{tikzpicture}[baseline={([yshift=-1ex]current bounding box.center)}]
    \begin{feynman}
    
        % Worldline through blob
        \vertex (wtl) at (-.8, 0);
        \vertex (wtr) at (.8, 0);
        \draw[thick, dotted] (wtl) -- (wtr);

        % Blob
        \node[draw, circle, fill=gray!30, minimum size=.4cm] (blob1) {};
        \node[draw, circle, fill=gray!30, minimum size=.4cm] (blob2) at (0, -.83) {};

        % Graviton leg (bottom)
        \vertex (gl) at (-.5, -1.4);
        \vertex (gr) at (.5, -1.4);
        \draw[thick, photon] (blob1) -- (blob2);
        \draw[thick, photon] (blob2.220) -- (gl);
        \draw[thick, photon] (blob2.320) -- (gr);
        
        % Labels
        \node at (-.7, -1.6) {$k_1$};
        \node at (.8, -1.6) {$k_2$};
        
    \end{feynman}
\end{tikzpicture}

%% file: diagrams/graviton_cut_2.tex
\begin{tikzpicture}[baseline={([yshift=-.5ex]current bounding box.center)}]
    \begin{feynman}
    
        % Worldline through blob
        \vertex (wtl) at (-.8, 0);
        \vertex (wtr) at (.8, 0);
        \draw[thick, dotted] (wtl) -- (wtr);

        % Blob
        \node[draw, circle, fill=gray!30, minimum size=.4cm] (blob1) {};
        \node[draw, circle, fill=gray!30, minimum size=.4cm] (blob2) at (0, -.83) {};

        % Graviton leg (bottom)
        \vertex (gl) at (-.5, -1.3);
        \vertex (gc) at (0, -1.4);
        \vertex (gr) at (.5, -1.3);
        \draw[thick, photon] (blob1) -- (blob2);
        \draw[thick, photon] (blob2) -- (gl);
        \draw[thick, photon] (blob2) -- (gc);
        \draw[thick, photon] (blob2) -- (gr);
        
        % Labels
        \node at (-.7, -1.4) {$k_1$};
        \node at (0, -1.65) {$k_2$};
        \node at (.8, -1.4) {$k_3$};
        
    \end{feynman}
\end{tikzpicture}

%% file: sections/07-conclusion.tex
\section{Conclusion}

    In this work we have developed a set of on-shell recursion relations for rational amplitude in WQFT.
    The principal challenge --- the poor large-$z$ behavior of an all-line shift --- was overcome through knowledge of the infrared structure of the theory.
    This was made possible thanks to a new soft graviton theorem~\eqref{eq:soft-graviton-theorem}, derived from Feynman rules and gauge invariance.
    It generalizes the known soft theorem of pure gravity by introducing new worldline contributions.

    Our main result, equation~\eqref{eq:recursion}, relates any rational amplitude with $n \geq 3$ graviton, to smaller amplitudes from both WQFT and pure gravity.
    A separate formula was given for the case $n=2$.
    These recursion relations decompose amplitudes into graviton and worldline-factorization channels, supplemented by soft residues, whose coefficients are completely determined by the soft theorems of Sec.~\ref{sec:soft-theorems}.
    Using the soft-worldline theorem~\eqref{eq:leading_soft_z}, any on-shell amplitude with external worldline fluctuations can be obtained from the corresponding all-graviton amplitude.
    Hence the recursion, combined with the soft-worldline theorem is sufficient to construct any rational amplitude of the theory.
    These can then be used as input for generalized unitarity methods to obtain integrands of more complicated amplitudes \cite{Haddad2025,He2025}.
    
    We checked the output of our formulas by an explicit computation of the two- and three-graviton amplitudes.
    The full $d$-dimensional results were reduced to compact spinor-helicity expressions given in Sec.~\ref{sec:results}.

    The most immediate extension would be the computation of higher-point amplitudes to provide further checks of the formula.
    Spin effects can be added through higher-dimensional worldline operators~\cite{Jakobsen2022a,Jakobsen2022b,Jakobsen2023}, and given their relevance to gravitational waves astronomy, this identifies a natural and physically motivated generalization of the present framework.
    Finally it may be worthwhile to investigate the possibility of two-lines, BCFW-style recursion.
    Amplitudes obtained this way would be expressed in spinor-helicity variables and be four-dimensional by nature but we could hope for much more compact expressions.

    Together with~\cite{He2025}, the present work shows that the full on-shell scattering amplitudes machinery can be successfully applied to worldline theories. 
    This bodes well for future high-precision calculations in gravitational-wave physics, where efficient amplitude-based techniques become increasingly relevant.

%% file: appendices/a-soft-z-theorems.tex
\section{Derivation of the soft-worldline theorems}
\label{app:soft_z_derivations}

    In this section we give derivations for both the leading and subleading soft-$z$ theorems, following closely Ref.~\cite{He2025}.

    \subsection{Leading soft-worldline theorem}

        The worldline Lagrangian in \eqref{eq:wl-action} is invariant under the spurionic transformation
        \begin{equation}
            b^\mu \to b^\mu + a^\mu, \qquad \delta x^\mu(\tau) \to \delta x^\mu(\tau) - a^\mu,
        \end{equation}
        for a constant vector $a$. 
        This directly implies the operator equation
        \begin{equation}\label{eq:operator_leading}
            \frac{\partial \mathcal{L}_{\rm wl}}{\partial b^\mu} 
            = \frac{\partial \mathcal{L}_{\rm wl}}{\partial (\delta x^\mu)}.
        \end{equation}

        Let $|\alpha\rangle$ be an arbitrary state. We define its overlap with an operator $\mathcal{O}(\tau)$ as the limit\footnote{To be precise, the limit should be taken while keeping the total energy $\omega + \sum_{i=1}^n uk_i + \sum_{j=1}^m \omega_j = 0$, where $i$ and $j$ run over the external gravitons and worldline fluctuations of the state $\alpha$. Without this precaution the expression would vanish due to the global energy-conservation delta distribution that appears when evaluating the LSZ formula.} of the Fourier transform
        \begin{equation}\label{eq:overlap}
            \langle O \rangle 
            = \lim_{\omega \to 0} \int d\tau \, e^{i\omega\tau} 
              \langle \alpha | O(\tau) | \Omega \rangle,
        \end{equation}
        where $|\Omega \rangle$ is the vacuum of the theory. 
        Since the free worldline Lagrangian is independent 
        of the impact parameter $b$, and $i\langle \mathcal{L}^{\rm wl}_{\rm int}\rangle = \mathcal{A}(\alpha)$, the overlap of the left-hand side of \eqref{eq:operator_leading} gives
        \begin{equation}\label{eq:lhs_leading}
            \left\langle \frac{\partial \mathcal{L}_{\rm wl}}{\partial b^\mu} \right\rangle 
            = -i \frac{\partial}{\partial b^\mu} \mathcal{A}(\alpha).
        \end{equation}

        For the right-hand side, we use the Schwinger-Dyson equations to rewrite the overlap with Euler-Lagrange to find
        \begin{align}\label{eq:rhs_leading}
            \left\langle \frac{\partial \mathcal{L}_{\rm wl}}{\partial (\delta x^\mu)} \right\rangle 
            &= \left\langle \frac{d}{d\tau} \frac{\partial \mathcal{L}_{\rm wl}}{\partial (\delta \dot{x}^\mu)} \right\rangle \\
            &= -m\langle \delta\ddot{x}^\mu \rangle 
            - m \left\langle \frac{d}{d\tau}\left[
            \bigl(h_{\mu\alpha}(b+u\tau) +  \delta x^\gamma \partial_\gamma h_{\mu\alpha}(b+u\tau) + \cdots\bigr) (u^\alpha + \delta \dot{x}^\alpha)\right] \right\rangle.
        \end{align}
        The first term is precisely the LSZ formula with one additional soft external $z$ leg: the 
        Fourier transform and the two $\tau$-derivatives produce a factor of $\omega^2$ that cancels the external propagator, effectively amputating the diagram,
        \begin{equation}\label{eq:lsz_leading}
            m\langle \delta\ddot{x}^\mu \rangle = i \lim_{\omega \to 0} \mathcal{A}^\mu(\omega, \alpha),
        \end{equation}
        where $\mu$ is to be contracted with the soft-worldline polarization $\zeta^\mu$. 
        Note that before taking the limit, the extra external worldline fluctuation is off-shell; the limit then enforces the on-shell condition on $\omega$. 
        The second term is a total derivative; since the expression inside is regular at $\omega = 0$ (there are no $1/\omega$ or $1/\omega^2$ propagators), the additional factor of $\omega$ from the Fourier transform makes it vanish in the limit.

        Equating both sides of \eqref{eq:operator_leading} and using the fact that the only $b$-dependence of rational amplitudes is through a global phase $e^{ib \cdot k}$ (see Sec.~\ref{sec:worldline-qft}), we obtain the algebraic version of the leading soft-worldline theorem that is given in Eq.~\eqref{eq:leading_soft_z}.

    \subsection{Subleading soft-worldline theorem}

        The derivation of the subleading theorem is similar to the previous one. 
        This time the invariance of $\mathcal{L}^{\text{wl}}$ under the spurionic symmetry
        \begin{equation}
            u^\mu \to u^\mu + a^\mu, \qquad \delta x^\mu(\tau) \to \delta x^\mu(\tau) - a^\mu \tau.
        \end{equation}
        translates to an operator equation, of which we take the overlap with an on-shell state $\alpha$
        \begin{equation}\label{eq:opNLOsoftZ}
            \frac{\partial \mathcal{L}_{\rm wl}}{\partial u^\mu} = \tau \frac{\partial \mathcal{L}_{\rm wl}}{\partial (\delta x^\mu)} + \frac{\partial \mathcal{L}_{\rm wl}}{\partial (\delta \dot{x}^\mu)}
            \longrightarrow
            \left\langle \frac{\partial \mathcal{L}_{\rm wl}}{\partial u^\mu} \right\rangle = \left\langle \frac{d}{d\tau}\left( \tau \frac{\partial \mathcal{L}_{\rm wl}}{\partial (\delta \dot{x}^\mu)} \right) \right\rangle,
        \end{equation}
        where again, we have applied the Euler-Lagrange equations. 
        Now as we split the Lagrangian in its free and interacting part $\mathcal{L}_{\rm wl} = \mathcal{L}_0 + \mathcal{L}_{\text{int}}$, we find that
        \begin{equation}\label{eq:split_subleading}
            \left\langle \frac{\partial \mathcal{L}_0}{\partial u^\mu} \right\rangle = -m \langle \delta\dot{x}_\mu \rangle
            \qquad \text{and} \qquad 
            i \left \langle \frac{\partial \mathcal{L}_{\text{int}}}{\partial u^\mu} \right\rangle = \frac{\partial}{\partial u^\mu} A(\alpha).
        \end{equation}
        Although the first term is singular as $\omega \to 0$, the same singularity appears on the right-hand side of \eqref{eq:opNLOsoftZ} so the divergences cancel exactly. 
        Indeed, the part of the RHS that is linear in the fields reads 
        \begin{equation}
            -m \frac{d}{d\tau}(\tau(\delta \dot{x}_\mu + h_{\mu \alpha}(b + u \tau)u^\alpha))
                = -m \tau \delta \ddot{x}_\mu - m \delta \dot{x}_\mu - m \frac{d}{d\tau}(\tau h_{\mu \alpha}(b + u \tau))u^\alpha.
        \end{equation}
        Comparing the term $-m \langle \tau \ddot{z}_\mu \rangle$ to an LSZ reduction formula we see that
        \begin{equation}
            -m \langle \tau \delta \ddot{x}_\mu \rangle = - \lim_{\omega \rightarrow 0} \frac{\partial}{\partial \omega} A(\omega,\alpha).
        \end{equation}
        As in the case of the leading soft theorem it turns out that all the other terms on the RHS will vanish when $\omega = 0$.
        We are then left with the subleading soft-worldline theorem given in Eq.~\eqref{eq:subleading-soft-z}.

%% file: appendices/b-local-gauge-invariant-tensor-decomposition.tex
\section{Local gauge-invariant tensor decomposition}\label{app:local-gauge-invariant}

    In Sec.~\ref{sec:soft-graviton}, we have used the claim that a local, gauge-invariant symmetric tensor could always be expressed as a sum of terms of the form
    \begin{equation}
        \left[ (B_1 \! \cdot \! q) B_2 -  (B_2 \! \cdot \! q) B_1 \right] \otimes \left[ (C_1 \! \cdot \! q) C_2 -  (C_2 \! \cdot \! q) C_1 \right],
    \end{equation}
    where $B_{1,2}$ and $C_{1,2}$ are local expressions in $q$.
    We prove this fact here, starting with the simpler result, for a local, gauge-invariant vector.

    \subsection{Local gauge-invariant vector}

        Let $V$ a finite dimensional vector space, equipped with a non-degenerate symmetric bilinear form.
        We pick a basis $\{ e_{(\mu)} \}$ for $V$.
        Consider a vector $E^\mu(q)$, which is analytic around $q=0$ (locality) and that satisfies the transversality condition $q \cdot E(q) = 0$ for any $q$ (gauge invariance).
        We claim that $E$ is of the form
        \begin{equation}
            E(q) = \sum_{\mu} \left[ (B_1^{(\mu)} \! \! \cdot \! q) B_2^{(\mu)} -  (B_2^{(\mu)} \! \! \cdot \! q) B_1^{(\mu)} \right].
        \end{equation}

        Because $E$ is analytic, we can write it as a sum over homogeneous polynomials in $q$ of increasing degree $\sum_{d=0}^\infty E_{(d)}$.
        Then if the property holds for homogeneous polynomials of arbitrary degree, it will necessarily hold in the general case.
        We can therefore safely assume that $E$ is a homogeneous polynomial of degree $d$ in the components of $q$.
        Let us define the antisymmetric tensor
        \begin{equation}
            A^{\mu \nu}(q) = \frac{1}{d+1} \left( \frac{\partial E^\nu}{\partial q_\mu} - \frac{\partial E^\mu}{\partial q_\nu} \right).
        \end{equation}
        Notice that contraction with $q$ gives
        \begin{equation}
            q_\mu A^{\mu \nu} = \frac{1}{d+1} \left( q_\mu \frac{\partial E^\nu}{\partial q_\mu} - q_\mu \frac{\partial E^\mu}{\partial q_\nu} \right) = \frac{1}{d+1}(d E^\nu + E^\nu) = E^\nu,
        \end{equation}
        where for the first partial derivative we used Euler's theorem for homogeneous polynomials and for the second we used $ \partial_\mu (q \! \cdot \! E(q)) = 0$.
        Consider the expression 
        \begin{equation}
            \frac{1}{2} \sum_{\mu \nu} A^{\mu \nu}(q) \left[ (e_{(\mu)} \! \cdot \! q) e_{(\nu)} - (e_{(\nu)} \! \cdot \! q)e_{(\mu)} \right].
        \end{equation}
        It has the desired form and looking at its components, we find that it is in fact equal to our vector field $E$.

    \subsection{Symmetric tensor}

        Now let us focus on the case of interest : a symmetric tensor field $E^{\mu \nu}(q)$, again analytic around 0 and that satisfies $q_\mu E^{\mu \nu}(q) = q_\nu E^{\mu \nu}(q) = 0$ for any $q \in V$.

        Again it is enough to consider the case where the components of $E$ are homogeneous polynomials of degree $d$ in the components of $q$.
        Because the vectors $E^{(\nu)}$ with components $(E^{(\nu)})^\mu = E^{\mu \nu}$ satisfy the same properties as the vectors in the previous section, we can define the degree $d-1$ homogeneous tensor 
        \begin{equation}
            A^{\mu \nu, \rho} = \frac{1}{d + 1} \left( \frac{\partial E^{\nu \rho}}{\partial q_\mu} - \frac{\partial E^{\mu \rho}}{\partial q_\nu} \right)
        \end{equation}
        it is antisymmetric in its first two indices and similar to the previous case, we have $q_\mu A^{\mu \nu,\rho} = E^{\nu \rho}$.
        A direct computation shows that we also have
        \begin{equation}
            q_\rho A^{\mu \nu, \rho} = \frac{2}{d+1} E^{[\mu \nu]} = 0.
        \end{equation}
        this transversality condition on $A$ lets us iterate the strategy and we define
        \begin{equation}
            B^{\mu \nu, \rho \sigma} =\frac{1}{d} \left( \frac{\partial A^{\mu \nu, \sigma}}{\partial q_\rho} - \frac{\partial A^{\mu \nu, \rho}}{\partial q_\sigma} \right).
        \end{equation}
        This new tensor is antisymmetric in $\mu,\nu$ and $\rho,\sigma$ and using the same arguments as before we show that $q_\mu q_\rho B^{\mu \nu, \rho \sigma} =E^{\nu \sigma}$.
        Finally, by looking at it componentwise we see that 
        \begin{equation}
            E(q) = \frac{1}{4} \sum_{\mu \nu \rho \sigma} B^{\mu \nu, \rho \sigma}(q) \left[ (e_{(\mu)} \! \cdot \! q) e_{(\nu)} - (e_{(\nu)} \! \cdot \! q)e_{(\mu)} \right] \otimes \left[ (e_{(\rho)} \! \cdot \! q) e_{(\sigma)} - (e_{(\sigma)} \! \cdot \! q)e_{(\rho)} \right],
        \end{equation}
        completing our proof.
        
        In this article we are only interested in the case of a symmetric tensor, however the result is still true in general.
        The proof for an antisymmetric tensor can be done with minor modifications, and because for a generic tensor $E^{\mu \nu} = E^{(\mu \nu)} + E^{[\mu \nu]}$, the general proof follows immediately.

%% file: appendices/c-spinor-helicity.tex
\section{Spinor helicity formalism in WQFT}\label{app:spinor-helicity}

    In this section we mainly record the additional ingredients required by the spinor-helicity formalism specific to WQFT.
    A thorough introduction to the spinor helicity formalism can be found in \cite{Elvang2015, Cheung2017}.
    For explicit spinor algebra computations, we use $(\sigma^\mu)^{\alpha \dot{\alpha}} = (\delta^{\alpha\dot\alpha}, \Vec{\sigma}^{\alpha\dot\alpha})$ and $\sigma^\mu_{\dot\alpha\alpha} = (\delta_{\dot\alpha\alpha}, -\Vec{\sigma}_{\dot\alpha\alpha})$, where the matrix vectors $\Vec{\sigma}$ denote the usual Pauli matrices.
    We also define the quantities
    \begin{equation}
    \begin{aligned}
        \sigma^{\mu \nu} &= \frac{1}{4}(\sigma^\mu \Bar{\sigma}^\nu - \sigma^\nu \Bar{\sigma}^\mu) \\
        \Bar{\sigma}^{\mu \nu} &= \frac{1}{4}(\Bar{\sigma}^\mu \sigma^\nu - \Bar{\sigma}^\nu \sigma^\mu).
    \end{aligned}
    \end{equation}
    
    To convert a $d$-dimensional formula written in terms of Lorentz products into a spinor helicity expression, we use the standard dictionary
    \begin{equation}
    \begin{aligned}
        p_i \! \cdot \! p_j &= \frac{1}{2}\langle ij \rangle [ij], \qquad &
        e_i^+ \! \cdot  e_j^+ &= \frac{1}{2} \frac{[ij] \langle \eta_j \eta_i\rangle}{\langle \eta_i i \rangle \langle \eta_j j \rangle}, \\[6pt]
        e_i^+ \! \cdot p_j &= \frac{1}{2} \frac{\langle j \eta_i \rangle [ij]}{\langle\eta_i i\rangle }, \qquad &
        e_i^- \! \cdot  e_j^- &= \frac{1}{2} \frac{\langle ji \rangle [\tilde{\eta}_i \tilde{\eta}_j]}{[i \tilde{\eta}_i] [j \tilde{\eta}_j]}, \\[6pt]
        e_i^- \! \cdot p_j &= \frac{1}{2} \frac{\langle j i \rangle [\tilde{\eta}_i j]}{[i \tilde{\eta}_i ] }, \qquad &
        e_i^+ \! \cdot  e_j^- &= \frac{1}{2} \frac{\langle j \eta_i \rangle [i \tilde{\eta}_j]}{\langle \eta_i i \rangle [j \tilde{\eta}_j]},
    \end{aligned}
    \end{equation}
    which implies normalization $e^+ \! \cdot e^- = -1/2$ as in \cite{Cheung2017}.
    $\eta_i$, $\tilde\eta_i$ are arbitrary reference spinors subject to the conditions $\langle\eta_i\, i\rangle \neq 0$ and $[\tilde\eta_i\, i] \neq 0$ respectively.
    
    WQFT introduces a new kinematic object: the four-velocity $u^\mu$ of the worldline.
    Since it is timelike, its bispinor matrix $u^{\alpha\dot\alpha}$ has rank two and cannot be written as a single outer product of Weyl spinors.
    It is possible to express it as a sum of two Weyl spinor products but this would introduce a total of four spinors.
    For simplicity we choose to keep $u^{\alpha\dot\alpha} = u^\mu \sigma_\mu^{\alpha \dot{\alpha}}$, or equivalently $u_{\dot\alpha\alpha} = u^\mu (\bar\sigma_\mu)_{\dot\alpha\alpha}$ as matrices.
    
    % For simplicity we choose to keep $u^{\alpha\dot\alpha} = u^\mu \sigma_\mu^{\alpha \dot{\alpha}}$  as a matrix, with $\sigma_\mu^{\alpha \dot{\alpha}} = (\delta^{\alpha\dot\alpha}, \Vec{\sigma}^{\alpha\dot\alpha})$.
    % Equivalently we have $u_{\dot\alpha\alpha} = u^\mu \sigma_\mu^{\dot\alpha \alpha}$  as a matrix, with $\sigma_\mu^{\alpha \dot{\alpha}} = (\delta^{\alpha\dot\alpha}, \Vec{\sigma}^{\alpha\dot\alpha})$.
    
    In this prescription we have for example $u \! \cdot \! p_i = \frac{1}{2} u^{\alpha\dot\alpha} p_{i\,\dot\alpha\alpha} = \frac{1}{2}\langle i|u|i] = \frac{1}{2}[i|u|i\rangle$.
    To account for any term involving $u^\mu$, we thereby extend our dictionary with
    \begin{equation}
        u \! \cdot \! p_i = \frac{1}{2} \langle i \vert u \vert i ], \qquad
        u \! \cdot \! e_i^+ = \frac{1}{2} \frac{\langle \eta_i \vert u \vert i ]}{\langle \eta_i i\rangle}, \qquad
        u \! \cdot \! e_i^- = \frac{1}{2} \frac{\langle i \vert u \vert \tilde{\eta}_i ]}{[ i \tilde{\eta}_i ]}.
    \end{equation}

    From there manipulation of SH variables in WQFT is mostly the same as in the general case, with an important addition regarding Schouten identities like
    \begin{equation}\label{eq:schouten}
        \langle a b \rangle \langle c d \rangle + \langle a c \rangle \langle d b \rangle + \langle a d \rangle \langle b c \rangle = 0.
    \end{equation}
    Because an expression $u \vert x ]$ really behaves like an angle spinor $y \rangle = u \vert x ]$, any angle spinor in \eqref{eq:schouten} can be replaced with some $u \vert x ]$.
    Note that if $y \rangle = u \vert x ]$, then $\langle y = - [x \vert u$ as can be shown by an explicit computation using spinor indices.
    This lets us derive new identities like
    \begin{equation}
        - u^2[1 2] \langle 12 \rangle - [1 \vert u \vert1 \rangle [2 \vert u \vert 2 \rangle + [ 1 \vert u \vert 2 \rangle [2 \vert u \vert 1 \rangle = 0,
    \end{equation}
    where we pulled the $u^2$ outside of the square brackets in the first term.
    We are allowed to do so thanks to the identity
    \begin{equation}
        u_{\dot{\alpha} \alpha} u^{\alpha \dot{\beta}} = \delta_{\dot{\alpha}}^{\dot{\beta}} u^2 + 2u_\mu u_\nu {(\Bar{\sigma}^{\mu \nu})_{\dot{\alpha}}}^{\dot{\beta}} = \delta_{\dot{\alpha}}^{\dot{\beta}} u^2,
    \end{equation}
    where the second term vanishes because $\Bar{\sigma}^{\mu \nu}$ is antisymmetric.
    
    This way we obtain numerous Schouten identities that can be used to tremendously simplify amplitudes written in SH variables and ultimately obtain the form displayed in section \ref{sec:results}.